# On Steganography in Lost Audio Packets

Wojciech Mazurczyk, Józef Lubacz, Krzysztof Szczypiorski
Warsaw University of Technology, Institute of Telecommunications
Warsaw, Poland, 00-665, Nowowiejska 15/19

**Abstract.** The paper presents a new hidden data insertion procedure based on estimated probability of the remaining time of the call for steganographic method called LACK (Lost Audio PaCKets steganography). LACK provides hidden communication for real-time services like Voice over IP. The analytical results presented in this paper concern the influence of LACK's hidden data insertion procedures on the method's impact on quality of voice transmission and its resistance to steganalysis. The proposed hidden data insertion procedure is also compared to previous steganogram insertion approach based on estimated remaining average call duration.

## 1. Introduction

LACK (Lost Audio PaCKets steganography) is a steganographic method, which modifies both RTP [5] packets and their time dependencies and it is intended for a broad class of multimedia, real-time applications like IP telephony. The method utilizes the fact that for usual multimedia communication protocols like RTP (Real-Time Transport Protocol) excessively delayed packets are not used for reconstruction of transmitted data at the receiver, i.e. the packets are considered useless and discarded.

LACK can be characterised by the following features: steganographic bandwidth, undetectability and steganographic cost. Steganographic bandwidth describes how much secret data we are able to send using a particular method per time unit. Undetectability is defined as an inability to detect a steganogram inside certain carriers. The most popular way to detect a steganogram is to analyse statistical properties of the captured data and compare it to the typical properties of that carrier. Steganographic cost characterises the degree of degradation of the carrier caused by the steganogram insertion procedure. The steganographic cost depends on the type of the carrier, and if it becomes excessive, it leads to easy detection of the steganographic method. For example, if the method uses voice packets as a carrier for steganographic purposes in IP telephony, then the cost is expressed in conversation degradation. If the carrier is certain fields of the protocol header, then the cost is expressed as a potential loss in that protocol functionality, etc.

It should be emphasised that the hidden data insertion procedures introduced and analysed in this paper can be utilized by decent LACK users who use their own VoIP calls to exchange covert data, but also by intruders who are able to covertly send data using third party VoIP calls (e.g. in effect of earlier successful attacks by using trojans or worms or by distributing modified versions of a popular VoIP software [17, 18]). This is a usual trade-off requiring consideration in a broader steganography context which is beyond the scope of this paper.

In this paper, we investigate LACK (Lost Audio PaCKets steganography), which was originally proposed in [12] and studied in [15]. This paper is an extension and continuation of the previous work presented in [16].

The contributions of this paper are:

- Detailed analysis of the LACK performance issues and of dependence of the insertion procedure on estimated VoIP call quality (Sec. 3 and Sec. 4).
- Extension of the previously proposed hidden data insertion procedure based on estimated remaining average call duration by considering also influence of the estimated call quality (Sec. 5.2).
- Introduction of a new hidden data insertion procedure based on the estimated probability of the remaining time of the call (Sec. 5.3). Also for this procedure influence of the estimated call quality is considered. For both methods LACK performance results are presented.
- Comparison of the both presented procedures for steganogram insertion in LACK (Sec. 5.4).

The rest of the paper is structured as follows. In Section 2 the basics of LACK functioning and detection is presented. In Section 3 LACK performance issues involved in using the method are discussed. Section 4 investigates dependence of the hidden data insertion rate *IR(t)* on estimated call quality. In Section 5 two methods for determining *IR(t)* based on estimated call duration are presented, analysed and compared. Section 6 concludes our work and indicates possible future research.

## 2. LACK Basics

The detailed description of LACK functioning is as follows (see Fig. 1). At the transmitter, one packet is selected from the RTP stream and its voice payload is substituted with bits of the steganogram (1). Then selected audio packet is intentionally delayed before transmitting (2). If an excessively delayed packet reaches a receiver unaware of the steganographic procedure, it is discarded (3), because for unaware receivers the hidden data is "invisible". However, if the receiver knows about the hidden communication, then instead of deleting the packet the receiver extracts the payload (4). Because the payload of the intentionally delayed packets is used to transmit secret information to receivers aware of the procedure, so no extra packets are generated.

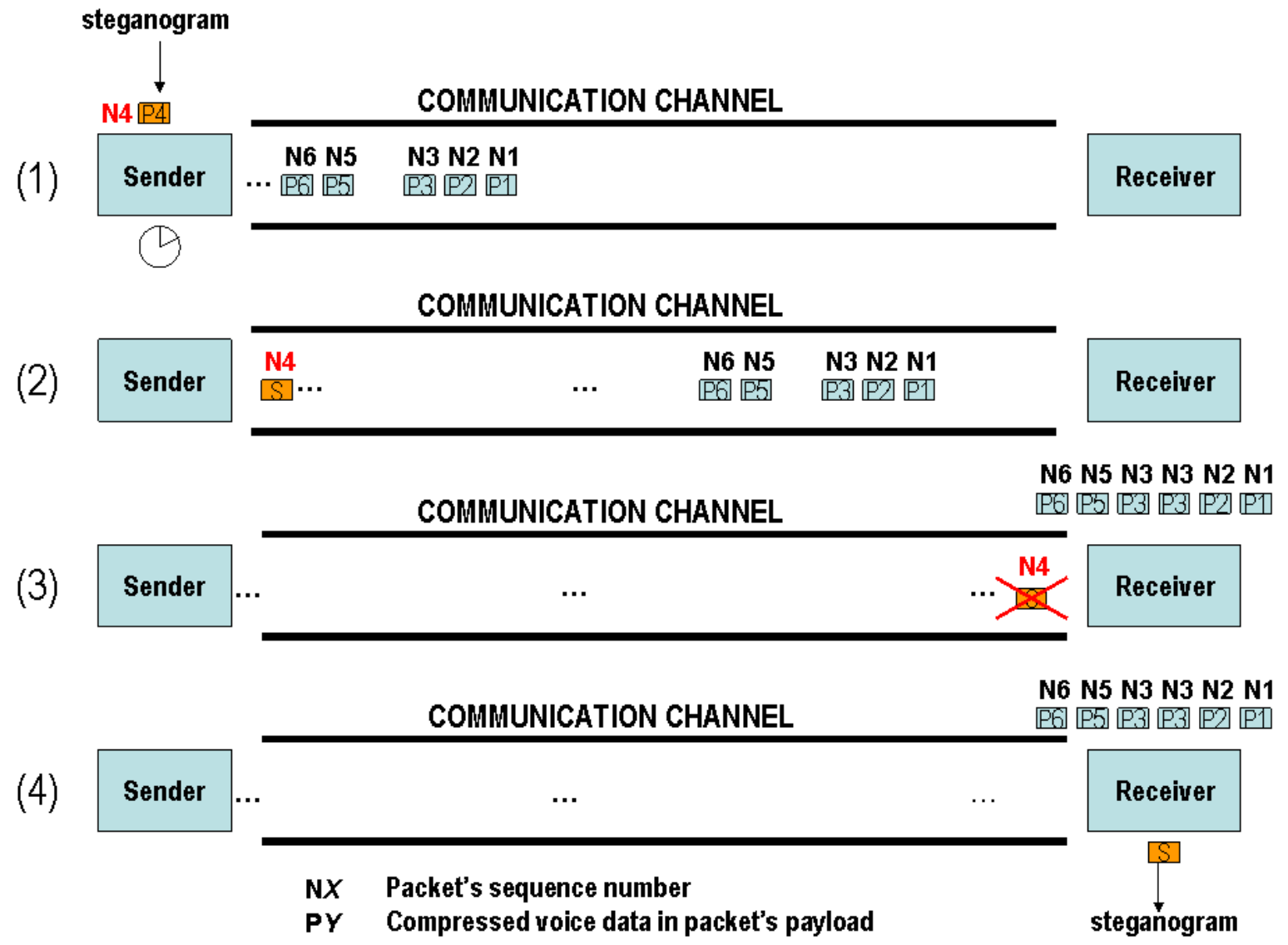

Fig. 1 The idea of LACK

LACK is a TCP/IP application layer steganography technique and is rather easy to implement. This is due to the fact that RTP is usually integrated in telephone endpoints (softphones) so access to RTP packets generation and modification is easier to perform than in the case of lower layer protocols like IP or UDP.

Steganalysis of LACK is hard to perform because packet loss in IP networks is a "natural phenomenon", so intentional losses introduced by LACK are not easy to detect, if kept on a reasonable level. Potential LACK steganalysis methods include:

- Statistical analysis of lost packets for calls in some sub-network. This type of steganalysis may be implemented with a passive warden [11] (or some other network node), based e.g. on information included in RTCP reports (cumulative number of packets lost field) exchanged between users during their communication or by observing RTP streams flow (packets' sequence numbers). If for some of the observed calls the number of lost packets is higher than average (or some chosen threshold) this may be used as an indication of potential use of LACK.
- Statistical analysis based on VoIP calls duration. If the call duration probability distribution for a certain sub-network is known, then statistical steganalysis may be performed to discover VoIP sources that do not fit to the distribution (the duration of LACK calls may be longer than non-LACK calls in effect of introducing steganographic data).
- An active warden [11] which analyses all RTP streams in the network (SSRC identifier and fields: Sequence Number and Timestamp from RTP header) can identify packets that are already too late to be used for voice reconstruction. The active warden may erase their payloads fields or simply drop them. A potential problem which arises in this case is to avoid eliminating delayed packets that still may be used for conversation reconstruction. The size of the jitter buffer at the receiver is, in principle, unknown to the active warden. If an active warden drops all delayed packets, then it will potentially drop packets that still can be useful for voice reconstruction. In effect, the quality of conversation may deteriorate considerably. Moreover, not only steganographic calls are affected but also non-steganographic ones are "punished".

## 3. LACK Performance Issues

The performance of LACK depends on many factors such as the details of the communication procedure (in particular the type of codec used, the size of the voice frame, the size of the receiving buffer, etc.) and on the network QoS (packet delay, packet loss probability and jitter). We discuss these issues in the following.

LACK's steganographic bandwidth and resistance to detection can be influenced by the following elements:

- The number of intentionally delayed RTP packets,
- The delay of the LACK packets,
- Network QoS – packet delay, packet loss probability and jitter,
- Features of the transmission devices – in particular type of the voice codec used (resistance to RTP packet losses and initial voice quality), the size of the RTP packet payload and the size of the jitter buffer.
- Hidden data insertion rate (*IR*) – number of bits of steganogram carried in a unit of time [bit/s].

In general, the more hidden information is inserted into the data stream, the greater the chance that it will be detected, e.g. by scanning the data flow or by some other steganalysis methods. Moreover, the more audio packets are used to send covert data, the greater the potential deterioration of the quality of VoIP connection. Thus the procedure of inserting hidden data has to be carefully chosen and controlled in order to minimize the chance of detecting inserted data and to avoid excessive deterioration of QoS. That is why the trade-off between achieved steganographic bandwidth, call quality deterioration and resistance to detection is always required.

Let assume that in a given moment of the call *t*, the packet is chosen from the RTP packets stream with probability $p_L(t)$ and network packet loss probability is $p_N(t)$. If $p_T$ denotes total acceptable probability of RTP packet losses then assuming independence of network packet losses from LACK choices we get

$$p_T \leq 1-(1-p_N(t))(1-p_L(t)), \qquad (2\text{-}1)$$

and in consequence

$$p_L(t) \leq \frac{p_T - p_N(t)}{1-p_N(t)}, \qquad (2\text{-}2)$$

which describes admissible level of the RTP packet losses introduced by LACK.

Exemplary relationships between probabilities $p_L(t)$, $p_N(t)$ and $p_T$ are illustrated in Fig. 2.

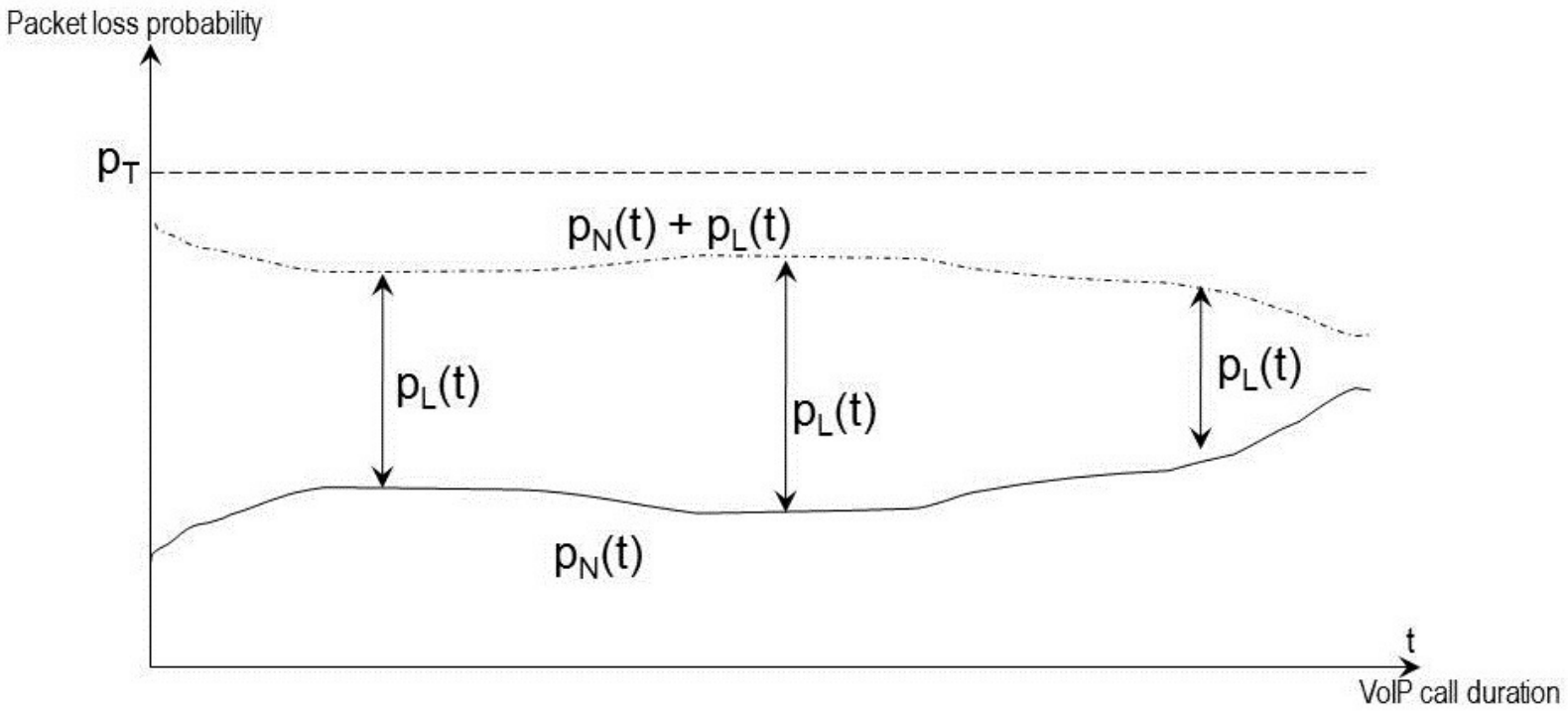

Fig. 2 LACK influence on total packet losses probability

For example, if $p_T = 0.05$ and $p_N(t_\xi) = 0.02$, then $p_L(t_\xi) \leq 0.03$.

To guarantee that an audio packet will be recognized as lost by receiver, it must be excessively delayed by the LACK procedure. To set this delay $d_L(t)$, the size of the receiver's jitter buffer must be taken into account. A jitter buffer is used to alleviate the jitter effect, i.e. the variations in packets arrival time caused by queuing, contention and serialization in the network. The size of the buffer is implementation-dependent. It may be fixed or adaptive, and is usually between 60 and 120 ms; RTP packet will be recognized as lost when the delay is greater than the delay introduced by the jitter buffer. LACK users have to exchange information about the sizes of their jitter buffers before starting the steganographic procedure. To limit the risk of detection of the hidden data, the delay chosen by LACK users should be as low as possible.

The RTP packet delay at the transmitter exit is equal

$$d_T(t) = d_D + d_K + d_E + d_L(t) \quad (2\text{-}3)$$

where:

$d_L(t)$ – intentional delay of RTP packet introduced by LACK,

$d_D$ – delay introduced by DSP (Digital Signal Processor) which depends on the type of the codec and is equal usually from 2 to 20 ms,

$d_K$ – delay introduced by voice coding (typically under 10 ms),

$d_E$ – delay caused by encapsulation (from 20 to 30 ms).

As mentioned above, the value of the intentional delay $d_L(t)$ introduced by LACK must be carefully chosen. Together with $d_N(t)$ introduced by network it must be greater than the size of the jitter buffer (Fig. 3), that is

$$d_T(t) + d_N(t) > t_B(t) \quad (2\text{-}4)$$

where:

$d_N(t)$ – delay introduced by network,

$t_B(t)$ – the size of the jitter buffer

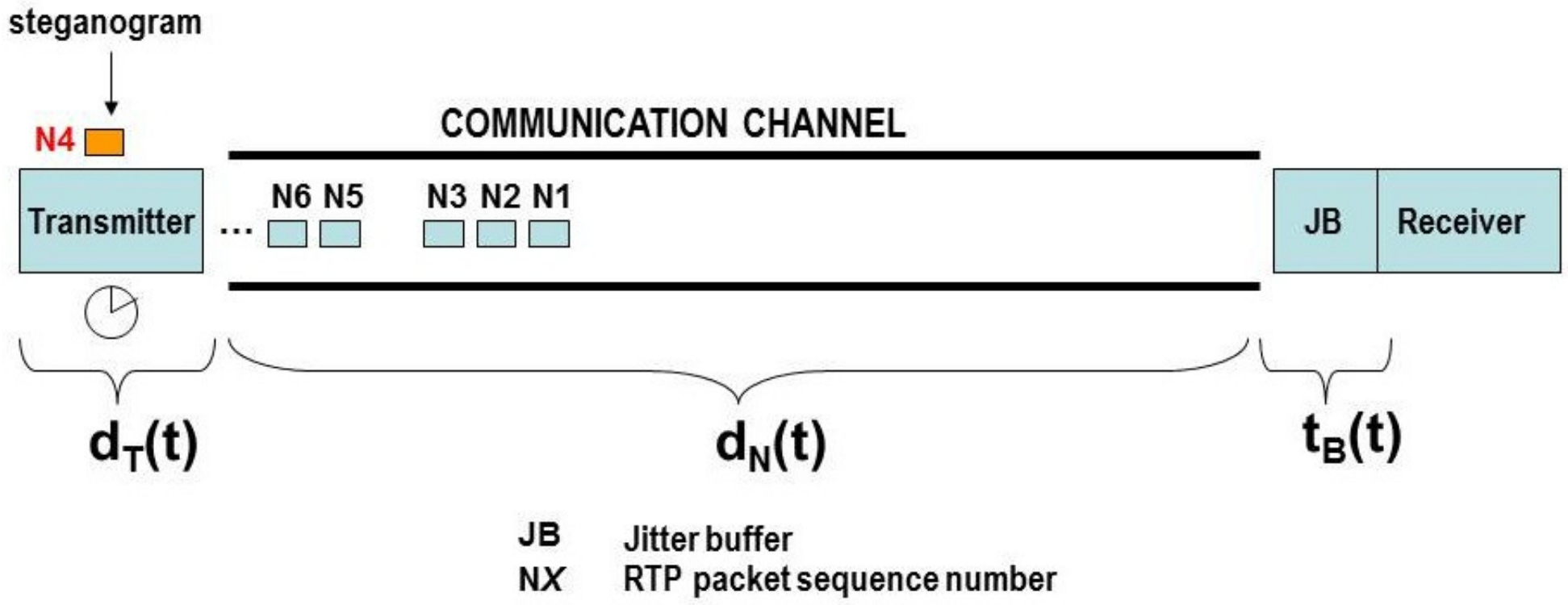

Fig. 3 Elements of LACK delay

The jitter buffer can be of a fixed size or adaptive. For example, if jitter buffer has a fixed size which is unchanged during the call and it does not consider network delay then delay at the transmitter output should be

$$d_T \geq t_B \quad (2\text{-}5)$$

and

$$d_L \geq t_B - d_D - d_K - d_E \quad (2\text{-}6)$$

Similar formulas can be derived for adaptive jitter buffer case.

Additionally, to ensure high steganographic bandwidth and undetectability of LACK it is necessary to observe network conditions while the call lasts. In particular packets losses, delay and jitter introduced by the network must be carefully monitored because they have influence on delay and packet losses that can be introduced by LACK without degrading perceived quality of the conversation. Because LACK uses legitimate RTP traffic, thus it increases overall packets losses. Thus, the level of the lost packets used for steganographic purposes must be controlled and dynamically adapted.

Information about network conditions during the call can be provided to the transmitter, for example, with use of SR (Sender Report), RR (Receiver Report) [5] or XR (Extended Report) [6] reports that are defined in RTCP protocol. If packet losses, delays and jitter are not monitored during the call, then they can be determined based on the historical, statistical data related to the network quality. However, it should be noted that RTP packet losses introduced by network can lead to lowering of the LACK steganographic bandwidth if the lost packet is a RTP packet that contains steganogram.

LACK steganographic bandwidth depends also on the codec used for VoIP conversation. Admissible level of packet losses usually is in range between 1 and 5%. For example, according to [20], maximum loss tolerance is 1% for G.723.1, 2% for G.729A and 3% for G.711 codecs. If a special mechanism to deal with lost packets at the receiver is utilized, e.g. the PLC (Packet Loss Concealment) [21], then the acceptable level of lost packets e.g. for G.711 codecs increases from 3% to 5%. The greater codec resistance to packet losses the better opportunity for achieving greater steganographic bandwidth for LACK. Thus the amount of steganographic data that can be inserted by LACK, and in effect the additional packet loss introduced by LACK, depends on the acceptable level of the total packet loss. For example, for the G.711 speech codec with data rate 64 kbit/s and data frame size of 20 ms, if the packet loss probability introduced by the LACK procedure is 0.5%, then the theoretical hidden communication rate is 320 b/s.

Another key element that influence LACK steganographic bandwidth and its resistance to steganalysis is hidden data insertion rate *IR(t)*, which is defined as a number of steganogram bits carried in a unit of time during the call [bit/s]. In general, the greater *IR(t)* the greater steganographic bandwidth and the greater degradation in voice quality and the easier steganalysis. *IR(t)* is influenced by:

- Assumed, acceptable call quality,
- Network conditions,
- The size of the steganogram,
- The duration of the call.

By applying correct procedure for determining *IR(t)* it is possible to control RTP packet losses and delays introduced by LACK without excessively affecting call quality and risking being detected. This aspect was carefully analysed in Sections 3 and 4.

In case if LACK is used sporadically by single user to transmit small amount of hidden data, utilizing complex methods for determining *IR(t)* is unnecessary because the chances of disclosure are very small and the effect on call quality is negligible. Complex variants of *IR(t)* calculation are important for such cases in which LACK is used frequently by single or a group of users in certain network localization.

In the simplest scenario *IR(t)* value can be fixed and constant during the call and calculated as *IR=S/T* where *S* is a size of the steganogram and *T* is predetermined duration of the call. Simple alternative is also possible by choosing constant *IR* and making the call last as long as the whole steganogram will be sent (the duration of the call is then equal *T=S/IR)*.The obvious disadvantage of such approach is however lack of relationship between *IR(t)* and voice quality and resistance to steganalysis.

*IR(t)* can be also set for the duration of the call based on statistical data (e.g. averages) on RTP packet losses and quality of the calls. However, it is not the proper solution for LACK, because it does not include potential changes in network conditions during the call and also the relationship between *IR(t)* and the size of the steganogram.

Methods for determining *IR(t)* based on current conversation quality, the size of the steganogram and duration of the call are considered in the following sections.

## 4. Dependence of the *IR(t)* on Estimated Call Quality

In this section we focus on the dependence of the insertion rate *IR* on estimated call quality resulting from packet loss. Call quality may be expressed in terms of subjective and objective quality measures. Objective measures are usually based on algorithms such as the E-Model [7], PAMS or PESQ [14]. The objective measures can be transformed into subjective quality measures. In our analysis we shall use the subjective measure MOS (Mean Opinion Score) [13] which according to [8] can be related to packet loss probability $p_N$ as follows

$$MOS_N(t) = \alpha \cdot \exp(\beta \cdot p_N(t)) + \gamma \qquad (3\text{-}1)$$

where $\alpha$, $\beta$ and $\gamma$ are network/service-type dependent parameters; for Skype telephony the parameters were evaluated to be [8]: $\alpha = 3.0829$, $\beta = -4.6446$ and $\gamma = 1.07$.

Since LACK introduces additional packet loss $p_L$ then in the above equation $p_N$ should be substituted with $p_N + p_L$

$$MOS_L(t) = \alpha \cdot \exp(\beta \cdot (p_N(t) + p_L(t))) + \gamma \qquad (3\text{-}2)$$

Fig. 4 shows the dependence of MOS on $p_N$ for different values of $p_L$ assuming $\alpha$, $\beta$ and $\gamma$ values estimated for Skype telephony.

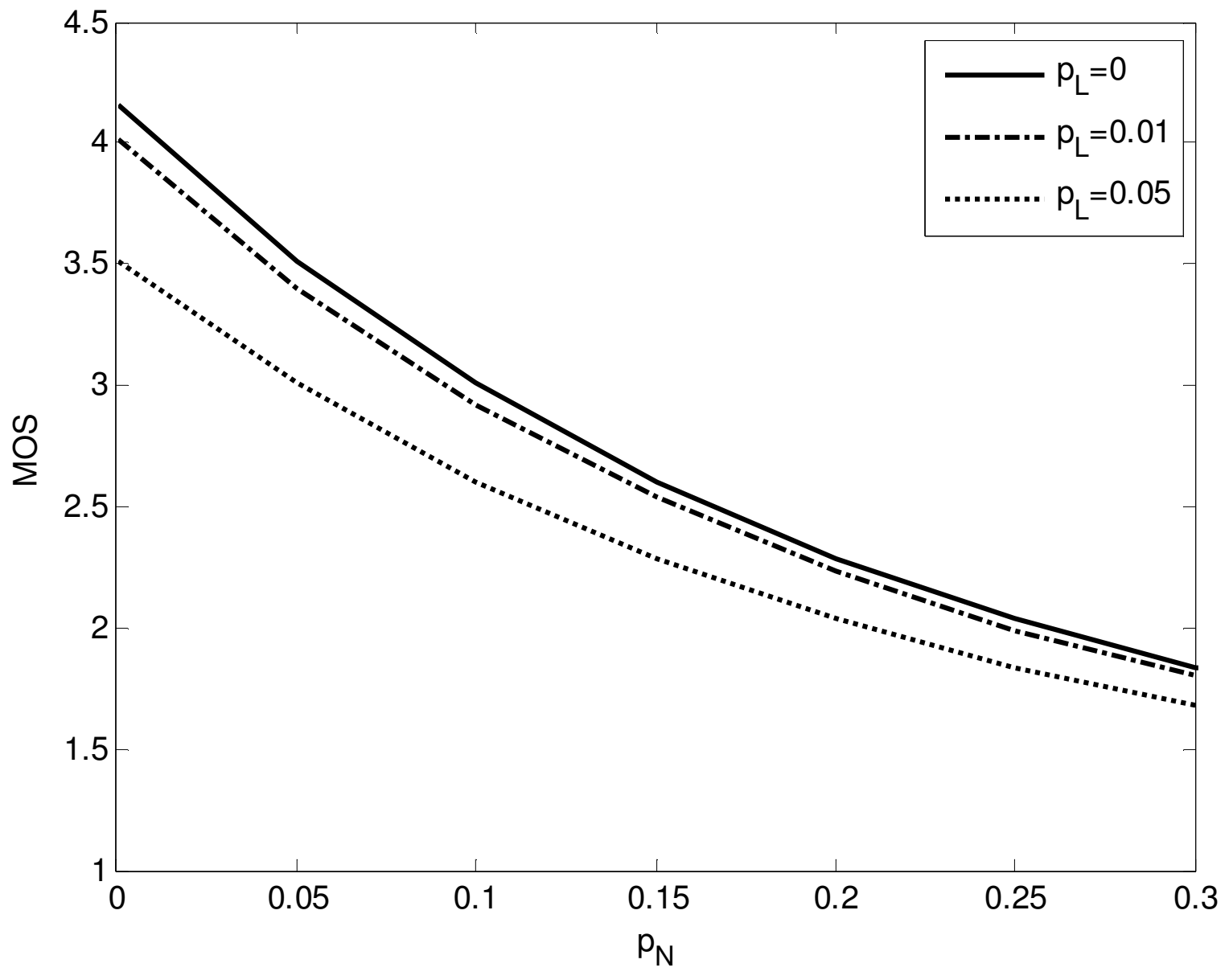

Fig. 4 MOS dependence on $p_N$ and $p_L$ for Skype telephony

The drop in call quality due to LACK utilization can be express as

$$\Delta MOS(t) = MOS_N(t) - MOS_L(t) = \alpha \cdot \exp(\beta \cdot p_N(t)) \cdot (1 - \exp(\beta \cdot p_L(t))) \qquad (3\text{-}3)$$

Let $IR_Q$ denote call quality dependent hidden data insertion rate expressed as MOS score. In general, $IR_Q$ can be:

- fixed during the VoIP call and determined based on historical, statistical data on calls quality or
- dynamically adjusted, while the call lasts, to the current estimation of voice quality

In the rest of this subsection we consider both cases described above.

### 4.1 Determining $IR_Q$ based on historical, statistical data on calls quality in given network

Let assume that the MOS probability distribution for a considered network in which LACK is to be used is known. Fig. 5 presents the MOS probability distribution for a VoIP network based on experimental data from [1].

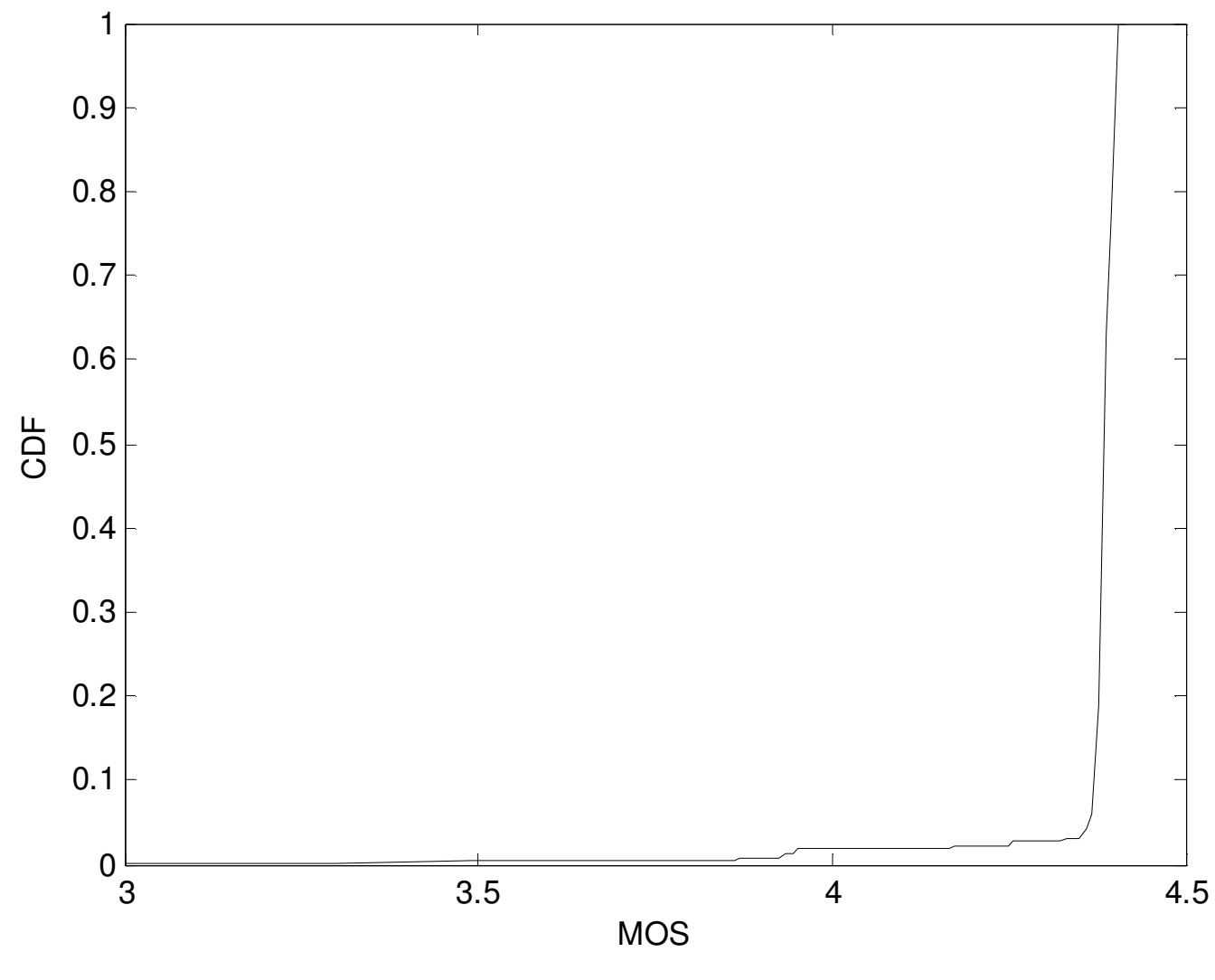

Fig. 5 MOS probability distribution (experimental data [1])

For given η the minimum, acceptable call quality MOS*

$$P(MOS > MOS^*) > \eta \tag{3-4}$$

Thus based on eq. 3-2 the upper limit of $p_L$ may be express as

$$p_L = \frac{\ln\left(\frac{MOS^* - \gamma}{\alpha}\right)}{\beta} - p_N \tag{3-5}$$

If $N_P$ is the number of RTP packets generated in a unit of time and $P_P$ is the length of a RTP packet data field (in bits), then

$$IR_Q \leq p_L \cdot N_P \cdot P_P \tag{3-6}$$

### 4.2 Determining *IR(t)* based on the current estimation of voice quality

An alternative to the approach described above is to adjust *IR(t)* based on online measurement of network parameters like: network losses, delays and jitter effect, which affect voice quality during the call. Such an approach would require online exchange of information on voice quality parameters between the sender and the receiver, e.g. with the use of the RTCP protocol (Sender Reports and Receiver Reports [5] or Extended Reports [6]). RTCP reports are exchanged by default every 5 seconds; however they can be sent more frequently if it is required (if network parameters change often). Based on this information estimated current voice quality is calculated $MOS_E(t)$.

For given upper limit of acceptable voice quality *MOS** while the call lasts it is verified whether $MOS(t) \geq MOS^*$. If this condition is fulfilled then

$$IR_Q(t) \leq N_P \cdot P_P \cdot \left( \frac{\ln\left(\frac{MOS_E(t) - \gamma}{\alpha}\right)}{\beta} - p_N(t) \right) \tag{3-7}$$

In any other case $IR_Q(t)=0$.

Dynamically adjusting $IR_Q(t)$ to current voice estimation can be troublesome and cause instabilities. Thus more practical approach is to utilize average values for given periods of time.

## 5. Dependence of the *IR* on Estimated Call Duration

In the following analysis we consider the dependence of the hidden data insertion rate *IR* for a particular call on the elapsed time of that call, i.e. we consider *IR* that is made time dependent. As shown in our analysis, such time-dependent *IR* procedure allows for decreasing the *IR* during the call duration, compared to the *IR* at call initiation time. In effect, the negative influence of LACK on QoS can be decreased and resistance to steganalysis increased, especially for call duration distributions with coefficient of variation much greater than 1. Available experimental data concerning VoIP call duration distributions seem to indicate that this is realistic for real-life VoIP calls. Our goal in this section is to express *IR* with the coefficient of variation for possibly wide range of call duration distributions.

### 5.1 VoIP call duration probability distribution

For PSTN the call duration probability distribution was well known due to extensive experimental research. For many decades the exponential distribution was assumed a good enough approximation for engineering purposes. VoIP is a relatively new service and thus only few reliable experimental data is available, so in many research papers concerning IP voice traffic (e.g. [2], [3], [4]) the exponential call duration is still assumed. Current experiments prove however that this assumption is far from being realistic.

Birke et al. [1] captured real VoIP traffic traces (about 150 000 calls) from FastWeb, an Italian telecom operator. The obtained call duration probability distribution is reproduced in Fig. 6 with a solid line. To illustrate qualitatively the degree in which the experimental results differ from exponential distribution it is drawn with broken line in Fig. 6. As can be seen, the differences are considerable and no straightforward approximation of the experimental data with standard distributions is available.

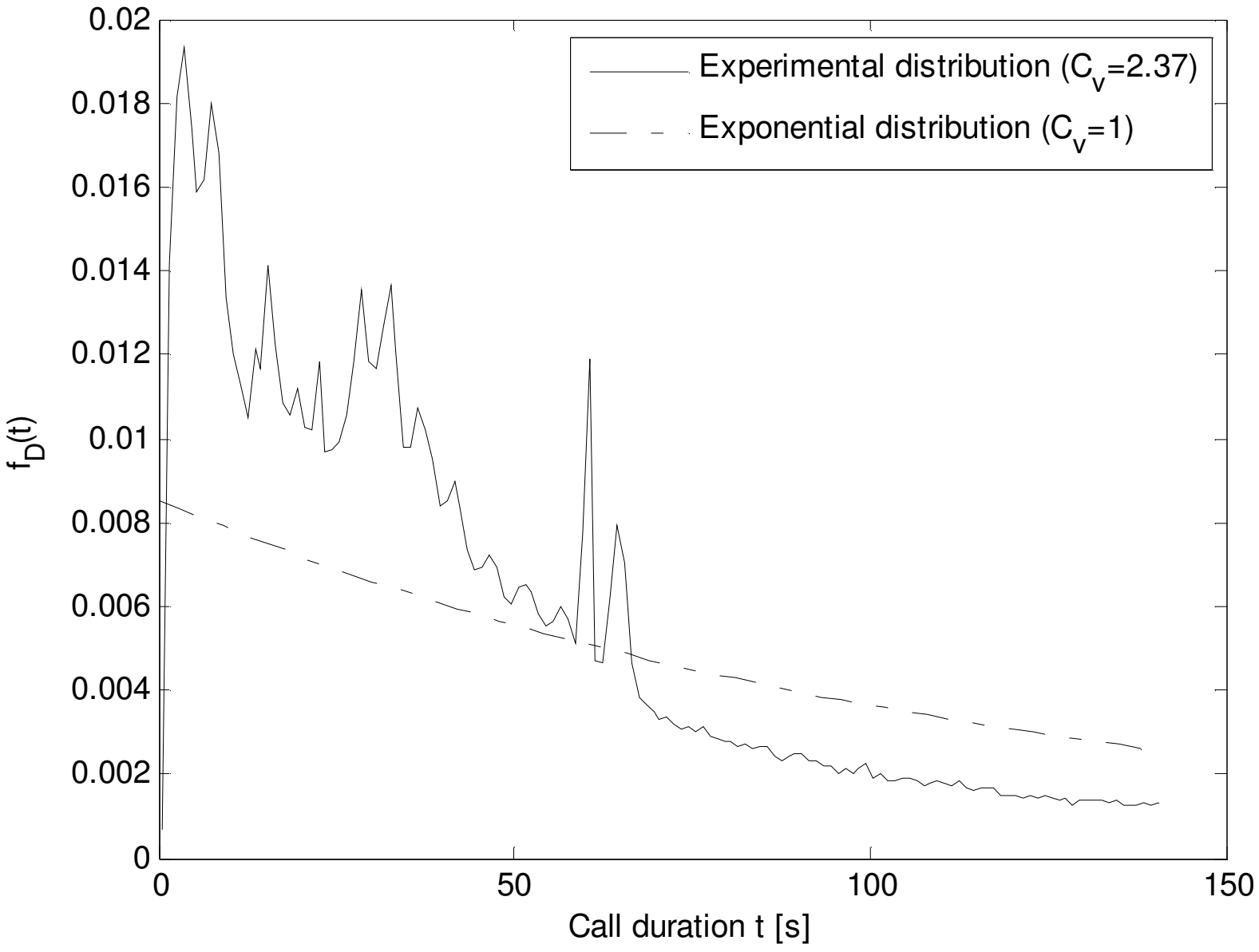

Fig. 6 VoIP call duration – comparison of experimental and exponential probability distributions

The experimental data from [1] yields average call duration *E(D) = 117.31 s* and standard deviation *σ(D) = 278.74*, thus the coefficient of variation $C_V = \sigma(D)/E(D) = 2.37$ (for the exponential distribution $C_V = 1$).

To achieve an analytic approximation of the experimental data a combination of some standard distributions can be used, for example:

$$f_D(t) = \begin{cases} \dfrac{1}{1.55t\sqrt{2\pi}} e^{-\frac{(\ln(t)-3.8)^2}{4.805}} & for \quad 0 \le t < 27.5 \\ 0.000114\text{e}^{-0.00114\text{ t}} + 0.027252\text{e}^{-0.03028\text{ t}} & for \quad 66.5 < t \le 27.5 \\ \dfrac{1}{1.55t\sqrt{2\pi}} e^{-\frac{(\ln(t)-3.8)^2}{4.805}} & for \quad 66.5 \le t \le 455 \end{cases} \qquad (4\text{-}1)$$

The above analytic approximation is quite complex and of little practical use for our purposes, i.e. for establishing the dependence of the insertion rate *IR* on some simple enough characterization of the call duration distribution.

Of course presented experimental data are not representative for IP telephony in general. However, it proves that for different applications of VoIP, including steganographic ones, the call duration probability distribution is far from exponential.

A reasonably wide range of call distribution types can however be achieved and effectively analysed/used with the 2-parameter Weibull distribution and appropriately chosen parameters: the shape parameter $k > 0$ and the scale parameter $\lambda > 0$. The complementary cumulative probability distribution function ($\overline{F}_D$) and probability density function ($f_D$) are as follows:

$$\overline{F}_D(t;k,\lambda) = e^{-\left(\frac{t}{\lambda}\right)^k}$$

$$f_D(t;k,\lambda) = \frac{k}{\lambda}\left(\frac{t}{\lambda}\right)^{k-1} e^{-\left(\frac{t}{\lambda}\right)^k} \qquad (4\text{-}2)$$

Average call duration and the coefficient of variation $C_V$ for this distribution are equal

$$E(D) = \lambda\Gamma\left(1+\frac{1}{k}\right)$$

$$C_V = \frac{\sqrt{\lambda^2\left[\Gamma\left(1+\frac{2}{k}\right)-\Gamma^2\left(1+\frac{1}{k}\right)\right]}}{\lambda\,\Gamma\left(1+\frac{1}{k}\right)} \tag{4-3}$$

The $\lambda$ parameter was set so to achieve the above experimental average call duration time $E(D) = 117.31$ and the $k$ parameter was varied so to obtain a wide range of $C_V$ values. In Tab. 1 the analysed values are summarized.

Table 1 Weibull distribution parameters $k$ and $\lambda$ and corresponding $C_V$ values

| **Weibull parameters** | **k=3.4, $\lambda$=130.57** | **k=2, $\lambda$=132.37** | **k=1.2, $\lambda$=124.71** | **k=1, $\lambda$=117.31** | **k=0.5, $\lambda$=58.65** |
|---|---|---|---|---|---|
| **$C_V$** | 0.32 | 0.52 | 0.84 | 1 | 2.23 |

In Fig. 7 the Weibull probability distribution is depicted for the parameters from Tab. 1 to illustrate the resulting wide range of distribution shapes. Note by the way that for $k = 1$ the Weibull distribution equals the exponential distribution ($C_V = 1$), for $k = 2$ it becomes the Rayleigh distribution ($C_V = 0.52$) and for $k = 3.4$ it resembles the normal distribution ($C_V = 0.32$).

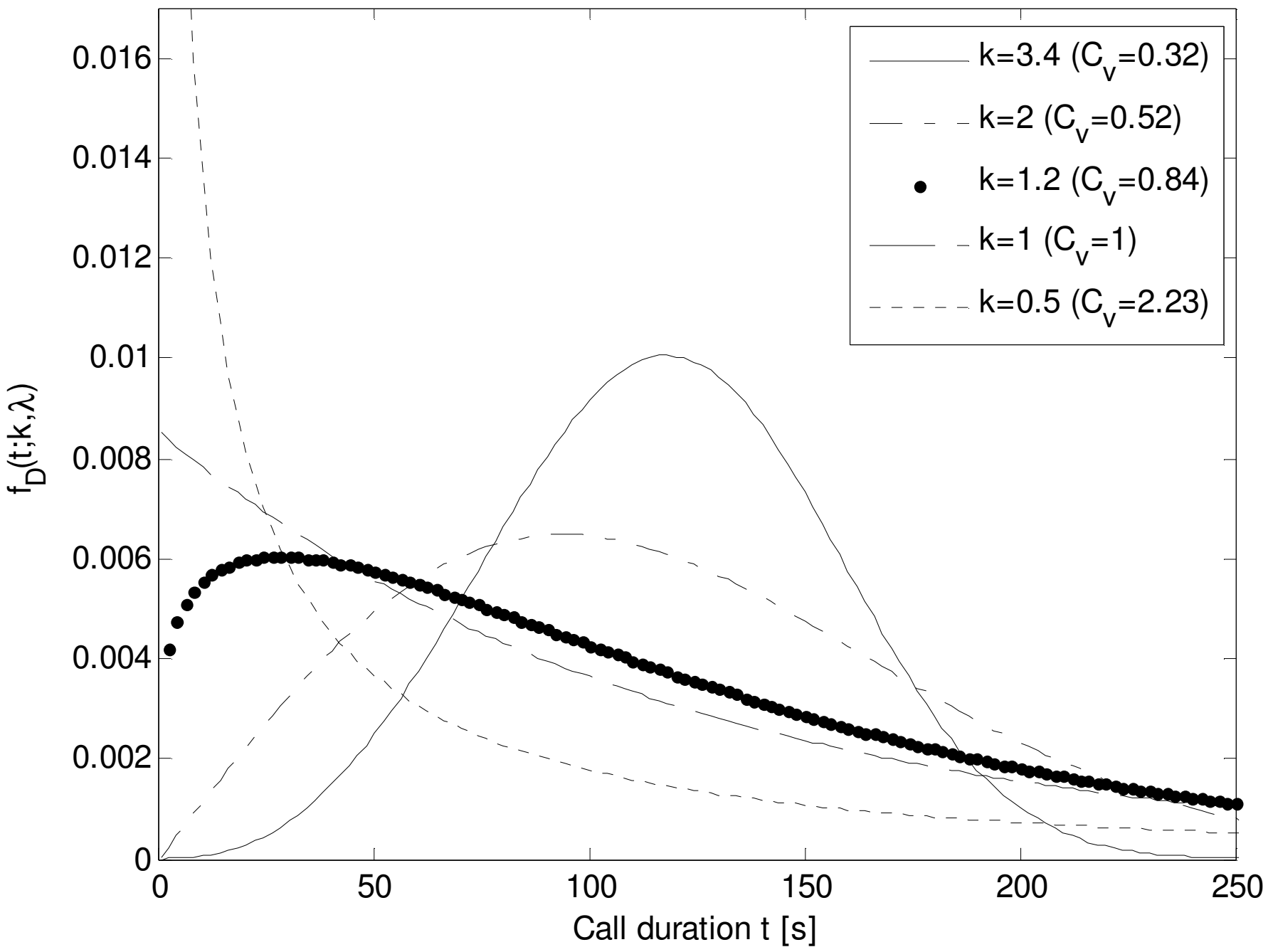

Fig. 7 Weibull distribution for various $k$, $\lambda$ and $C_V$

### 5.2 Dependence of *IR(t)* on estimated remaining average call duration

The following method of determining *IR(t)* was originally proposed in [16]. Here it is extended by considering also the call quality and analysed in more detail.

For an arbitrary instant of a call the average residual call duration is well known to be equal

$$E(R) = \frac{E(D^2)}{2E(D)} \tag{4-4}$$

or equivalently

$$E(R)=\frac{C_V^2+1}{2}E(D) \tag{4-5}$$

Suppose that at the beginning of a call the insertion rate is set to *IR(0) = S/E(D)*, where *S* is amount of data to be sent covertly. If $C_V>1$ then *E(R) > E(D)*, which seems to be the case for VoIP real-world calls as indicated above, then beginning from some arbitrary instant of the call we may decrease the insertion rate to *IR = S/E(R)*, which is beneficial from the point of view of call quality and resistance to detection of the hidden data.

The above indicates that it is reasonable to make the insertion rate dependent on the elapsed time of a call. It is nevertheless not practical to use the classical definition of residual call duration since it involves an arbitrary time instant and not the current call duration. We are rather interested in the expected call duration on condition it has already lasted *t* units of time:

$$E(D \mid D>t)=\frac{1}{P(D>t)}\int_t^{\infty} x f_D(x)dx = t+\frac{1}{\overline{F}_D(t)}\int_t^{\infty}\overline{F}_D(x)dx \tag{4-6}$$

for random variable *D* which values are from range [0, ∞). This leads to the following estimations

For *t=0 E(D|D>0) = E(D)*

For every *t*

$$E(D \mid D>t)\ge t$$

$$E(D \mid D>t)\ge E(D) \tag{4-7}$$

because

$$E(D \mid D>t)\ge \frac{1}{P(D>t)}\int_t^{\infty} t f_D(x)dx = t$$

and

$$\begin{aligned}E(D \mid D>t)&=t+\frac{1}{\overline{F}_D(t)}\int_t^{\infty}\overline{F}_D(x)dx \ge t+\int_t^{\infty}\overline{F}_D(x)dx = \\ &= t+\int_0^{\infty}\overline{F}_D(x)dx-\int_0^{t}\overline{F}_D(x)dx \ge E(D)\end{aligned} \tag{4-8}$$

It is worth noting that for exponential distribution *E(D|D>t) = t + E(D).*

Using above estimations it is possible to determine set of admissible values for *E(D|D>t)*, which is illustrated in Fig. 8.

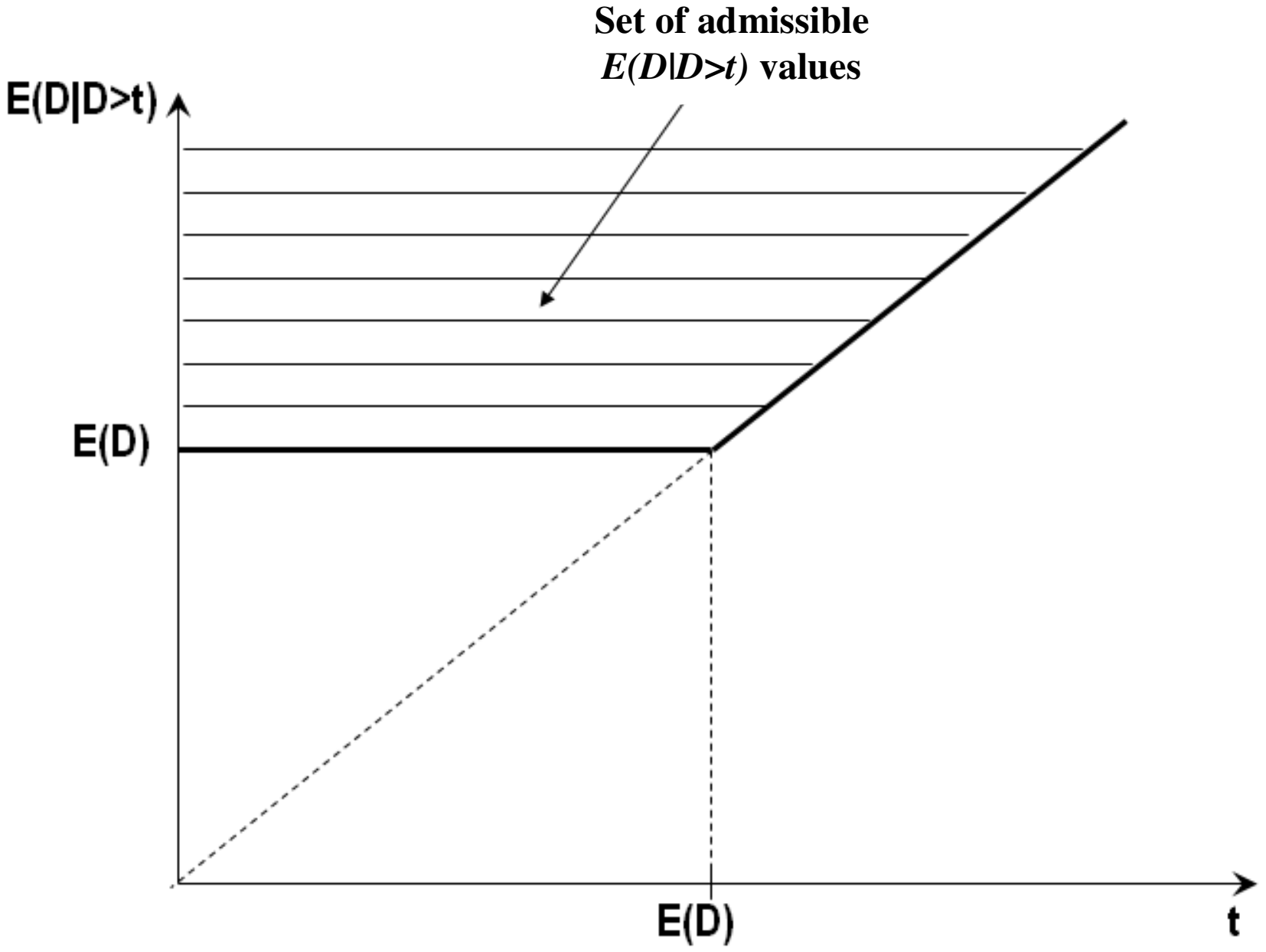

Fig. 8 Admissible values for *E(D|D>t)*

Upper limit of *E(D|D>t)* is as follows

$$E(D \mid D > t) = \frac{1}{P(D > t)}\left( E(D) - \int_0^t x f_D(x) dx \right) \le \frac{E(D)}{P(D > t)} \quad (4\text{-}9)$$

For 2-parameter Weibull distribution considered in Section 4.1

$$E(D \mid D > t) = t + e^{\left(\frac{t}{\lambda}\right)^k} \int_t^{\infty} e^{-\left(\frac{x}{\lambda}\right)^k} dx \quad (4\text{-}10)$$

and

$$E(D \mid D > t) \le e^{\left(\frac{x}{\lambda}\right)^k} \lambda \Gamma\left(1 + \frac{1}{k}\right)$$

$$E(D \mid D > t) \ge t \quad (4\text{-}11)$$

$$E(D \mid D > t) \ge \lambda\, \Gamma\left(1 + \frac{1}{k}\right)$$

For chosen parameters from Tab.1 we obtain results shown in Fig. 9. The figure shows also the *E(D|D>t)* function for the experimental data presented in Fig. 6.

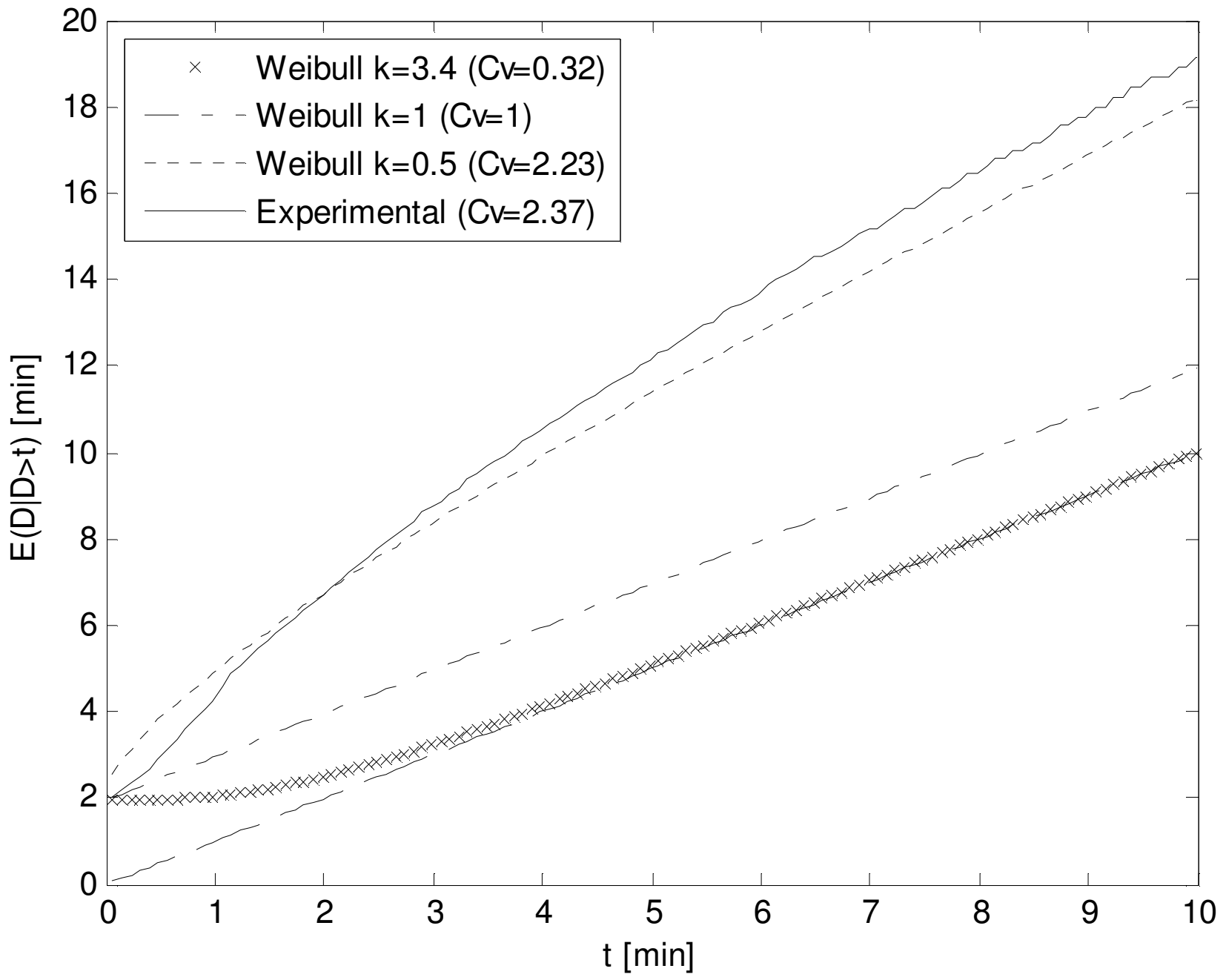

Fig. 9 *E(D|D>t)* for different Weibull distributions and for the experimental data distribution

The curves from Fig. 10 may be approximated with good accuracy as follows

$$E(D \mid D > t) \approx 1.32 C_v + t\sqrt{C_v} + 0.59 \quad [\text{min}] \quad (4\text{-}12)$$

If $S_R(t)$ is the amount of data remaining to be sent covertly at instant $t$ of the call

$$S_R(t) = S - \int_0^t IR(x) dx \quad (4\text{-}13)$$

then the insertion rate at time $t$ is

$$IR(t) = \begin{cases} \dfrac{S_R(t)}{E(D \mid D > t)} & for \quad IR(t) < IR_Q(t) \\ IR_Q(t) & for \quad IR(t) \ge IR_Q(t) \end{cases} \quad (4\text{-}14)$$

where $IR_Q(t)$ is calculated as described in Section 3.

Based on results presented in Fig. 9 and eq. 4-14, assuming $S = 1000$ bits, the *IR(t)* functions for chosen Weibull distributions are presented in Fig. 10. For the sake of simplicity we assumed that $IR(t)<IR_Q(t)$ i.e. no limitations related to call quality. These limitations are considered in Fig. 11.

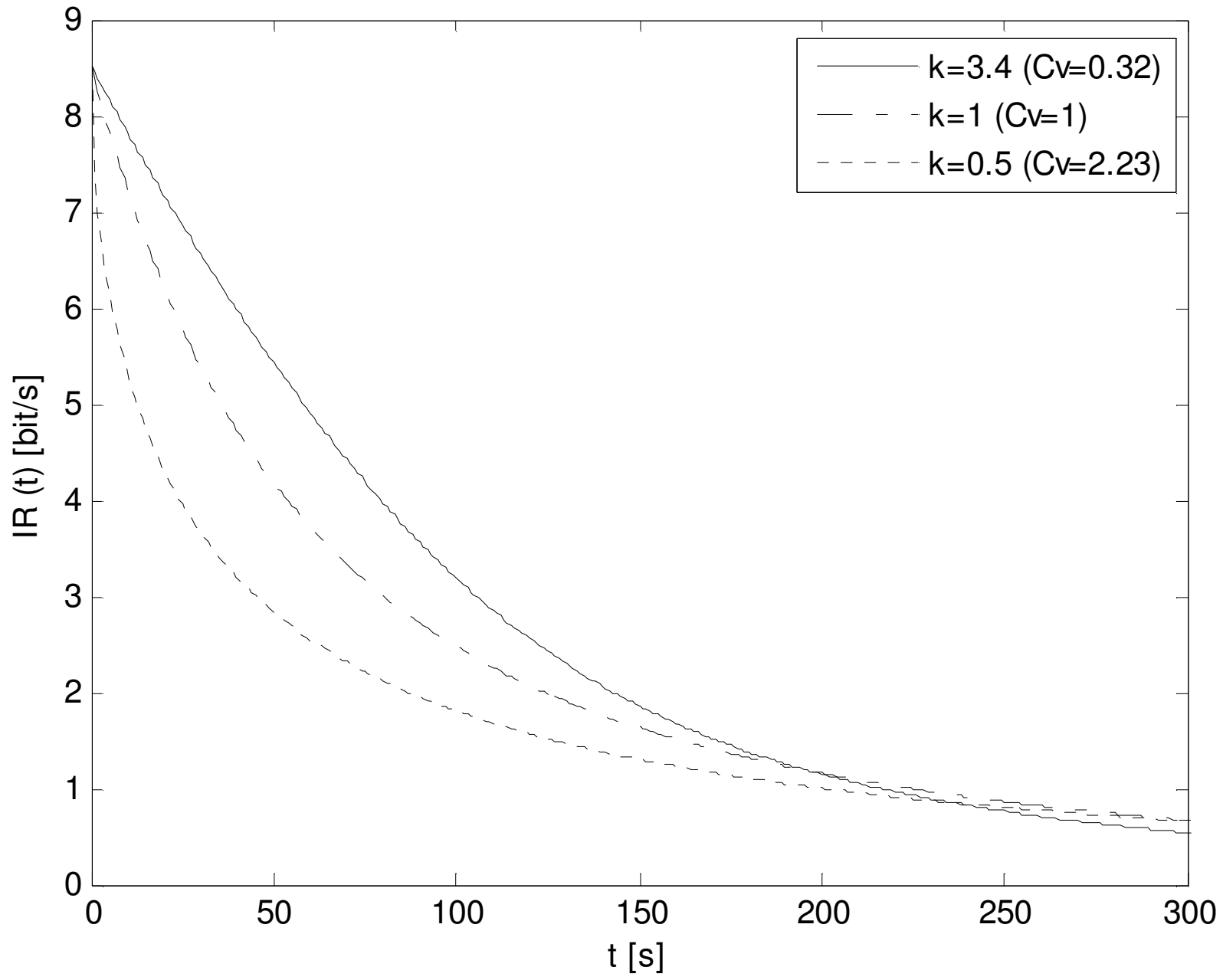

Fig. 10 *IR(t)* for chosen Weibull distributions, $S=1000$ bits ($IR(t)<IR_Q(t)$)

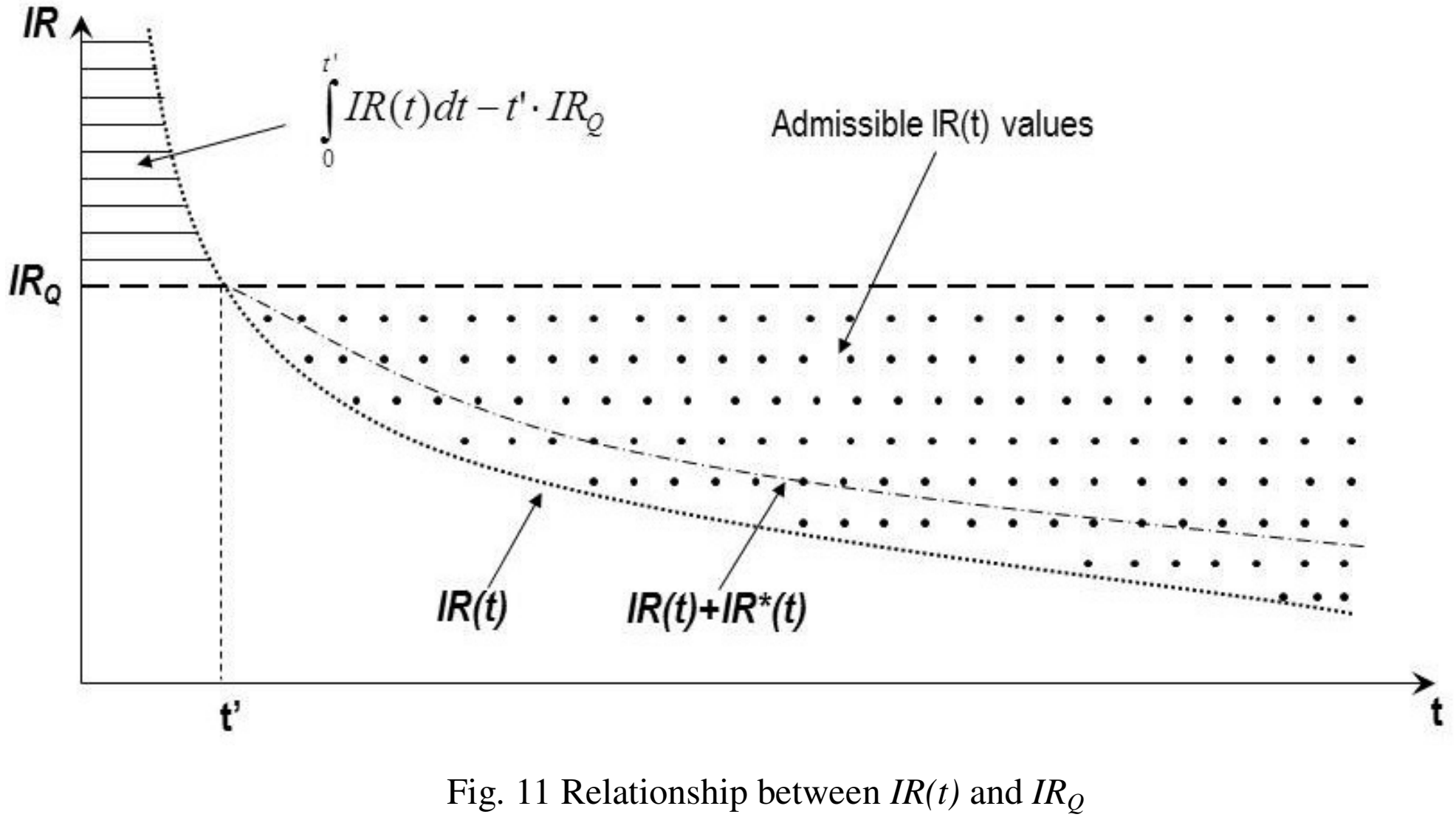

Fig. 11 Relationship between *IR(t)* and $IR_Q$

Consider that if $IR(t) > IR_Q$ for $t < t'$ then

$$\int_0^{t'} IR(t)dt - t' \cdot IR_Q \tag{4-15}$$

describes this part of the steganogram which will be sent if we do not consider the limitation $IR(t) < IR_Q(t)$ in range *[0, t')*. Such "arrear" can be aligned by increasing *IR(t)* for $t > t'$ (with limitation $IR(t) < IR_Q(t)$) this situation is illustrated in Fig. 12 with $IR(t) + IR^*(t)$ curve which can be for example expressed as

$$IR^*(t) = \frac{\int_0^{t'} IR(t)dt - t' \cdot IR_Q}{E(D \mid D > t')} \quad (4\text{-}16)$$

In Fig.12-14 dependence of *IR(t)* on steganogram size under limitation $IR(t) < IR_Q(t)$ is presented for given moments of VoIP call (for obtained results we assumed the same probability distributions and their parameters as in the previous calculations).

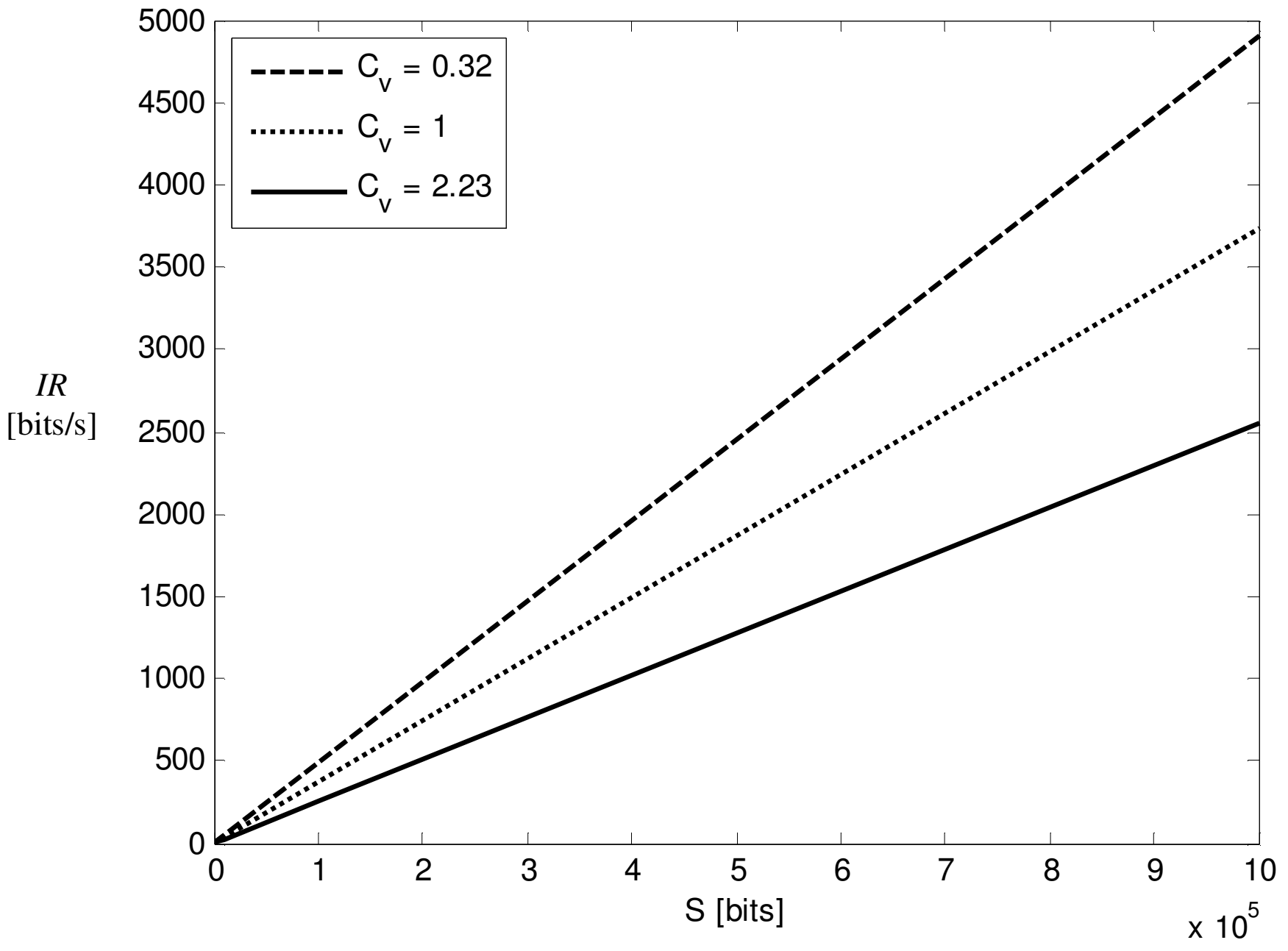

Fig. 12 Dependence of *IR(t)* on *S,* for $t = 60$ s and chosen $C_V$ values

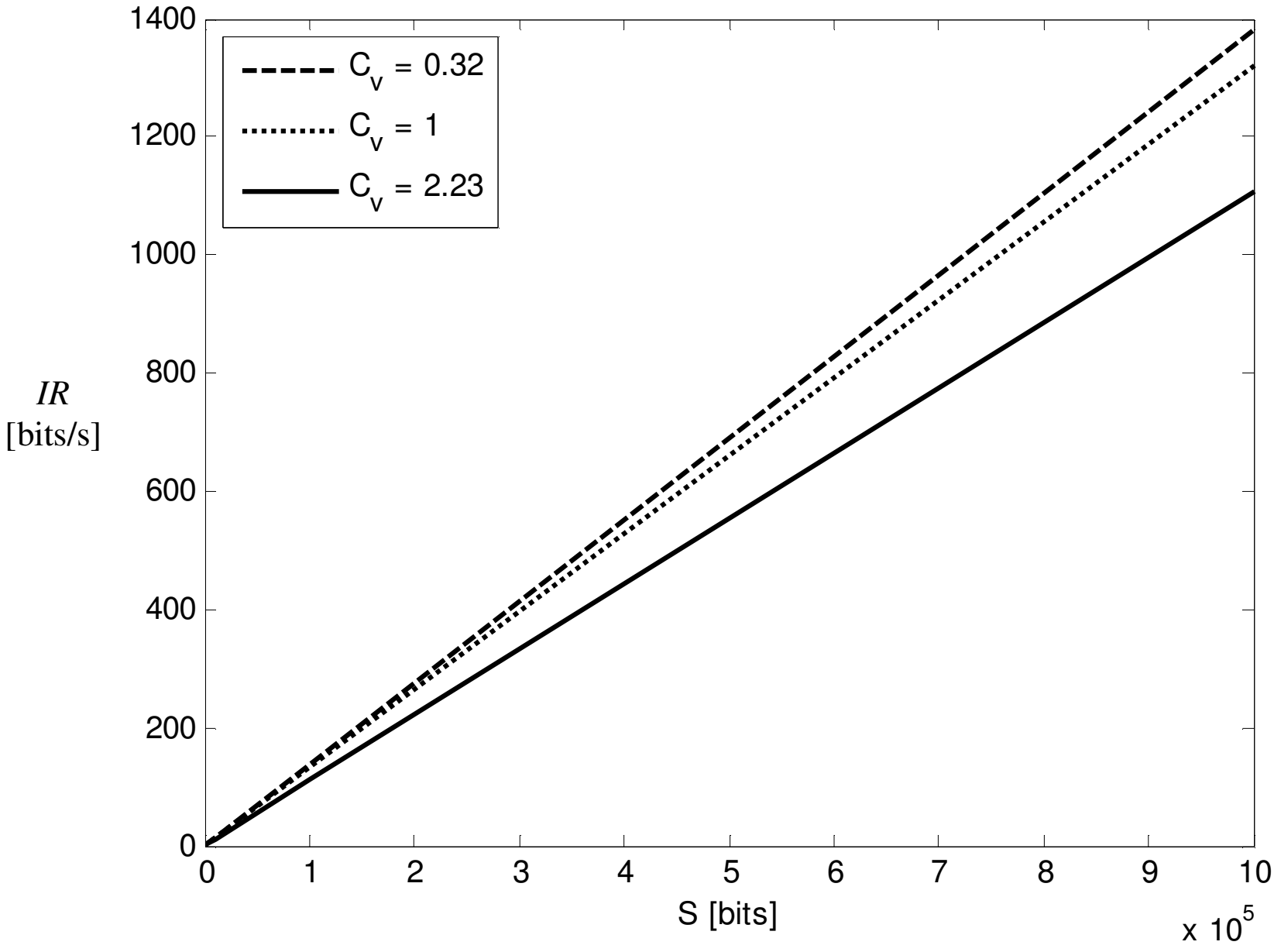

Fig. 13 Dependence of *IR(t)* on *S,* for $t = 180$ s and chosen $C_V$ values

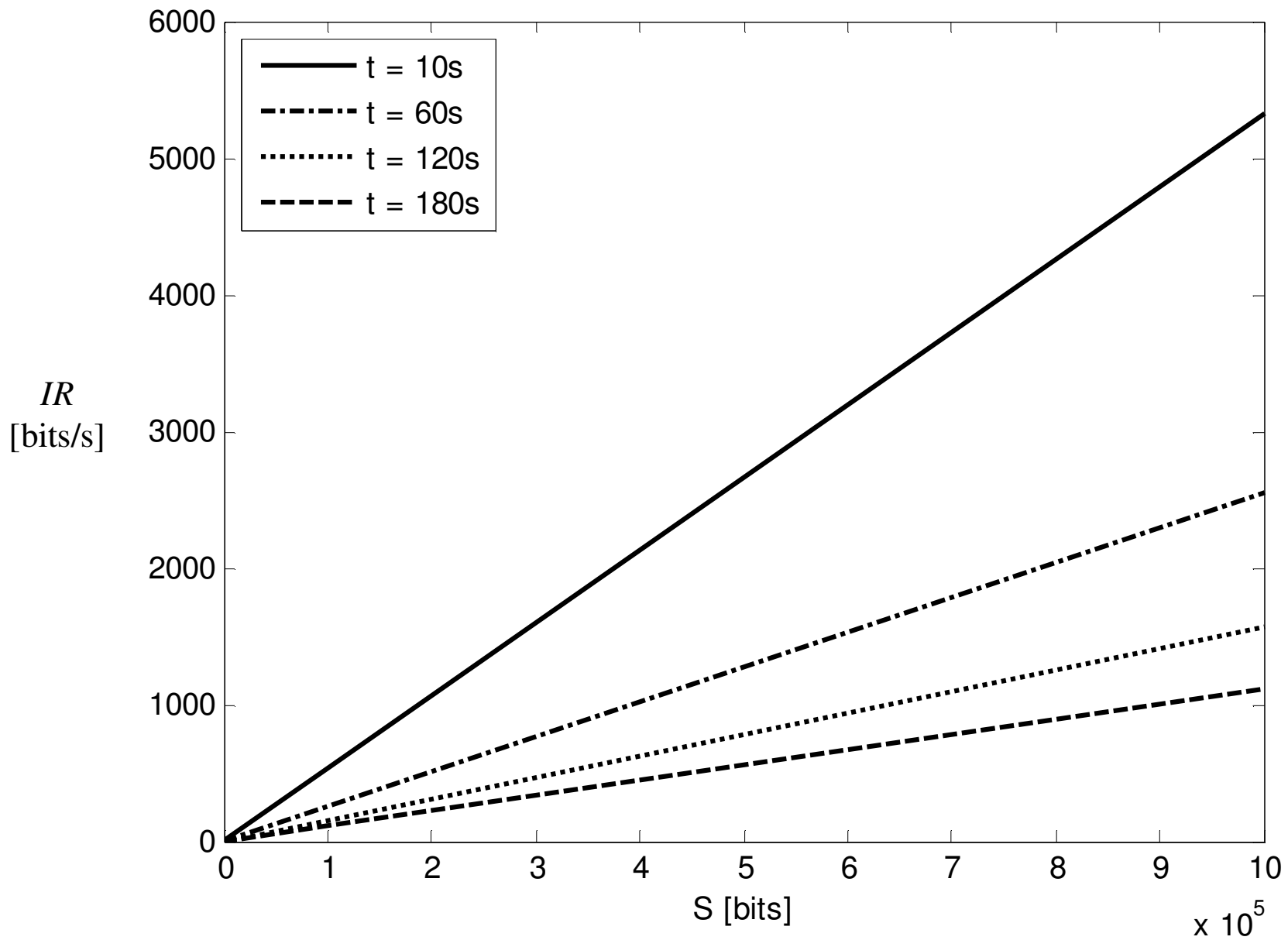

Fig. 14 Dependence of *IR(t)* on *S,* for chosen moments of VoIP call for $C_V=2.23$

The total effect – „gain" – from applying the procedure described above which relates *IR(t)* and *E(D/D>t)* and which results from decreasing *IR(t)* when compared to its initial value *IR(0)* is presented in Fig. 15. This is the desired effect which was aimed at: as the call proceeds, the *IR* is adjusted – decreased – according to the expected remaining duration of the call, which is, as already mentioned, beneficial from the point of view of voice quality and resistance to steganalysis. In quantitative terms the decrease in *IR(t)* – *X(t)* – is expressed by eq. 4-17 and total gain – *Z* – by eq. 4-18.

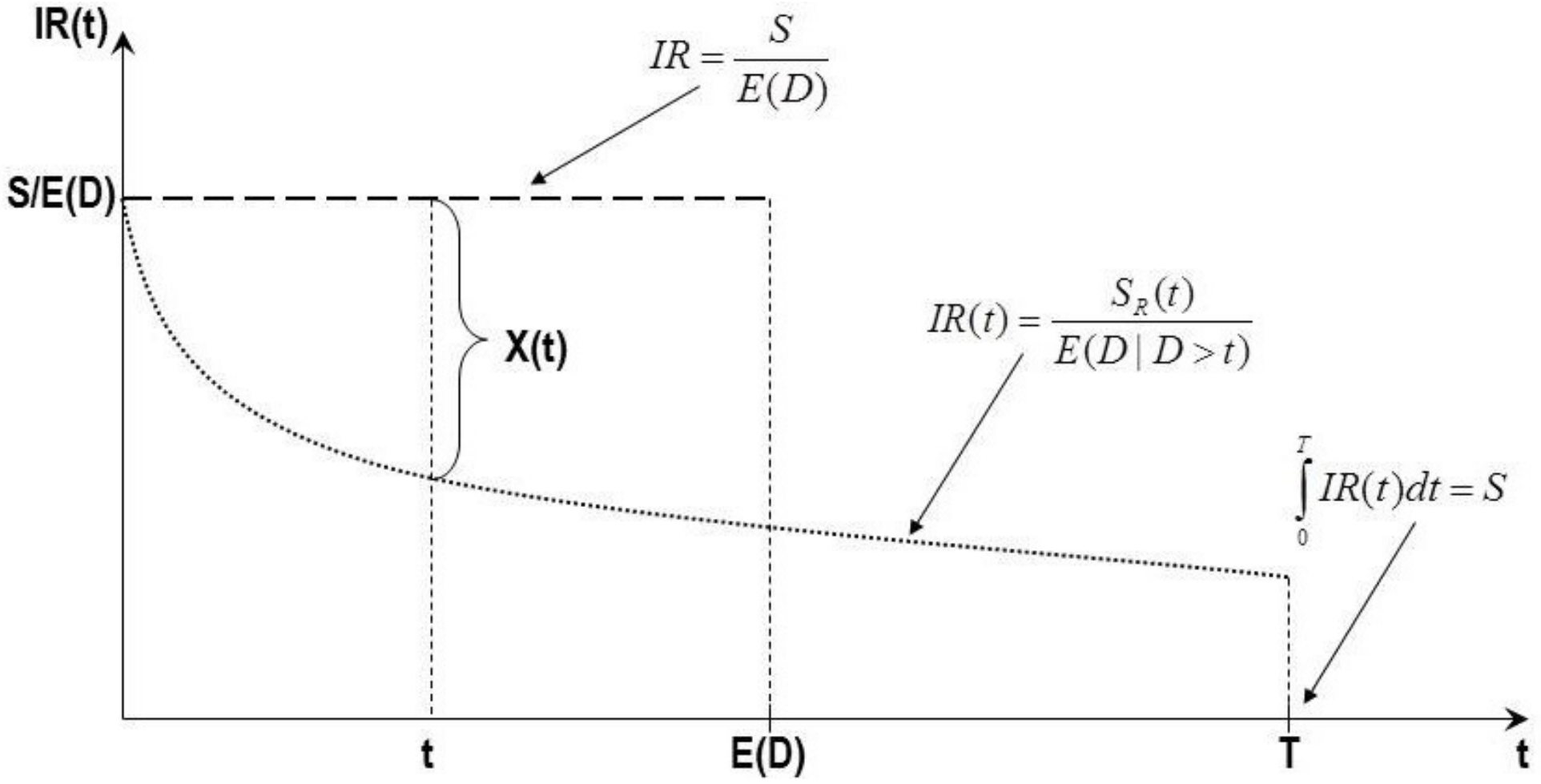

Fig. 15 The effect of using *IR(t)* based on *E(D|D>t)*

$$X(t) = IR(0) - IR(t) = \frac{S}{E(D)} - \frac{S - \int_0^t IR(x)dx}{E(D \mid D > t)} \tag{4-17}$$

$$Z = \int_0^T X(t)dt \tag{4-18}$$

*X(t)* can be also related to call quality expressed in MOS scale as follows. For fixed, constant *IR* = *S/E(D)* call quality can expressed as

$$MOS_{E(D)}(t) = \alpha \cdot \exp(\beta \cdot (p_N(t) + p_{E(D)})) + \gamma \qquad (4\text{-}19)$$

for case of dependence of *IR(t)* on *E(D|D>t)* it is

$$MOS_{E(D|D>t)}(t) = \alpha \cdot \exp(\beta \cdot (p_N(t) + p_{E(D|D>t)}(t))) + \gamma \qquad (4\text{-}20)$$

where $p_{E(D)}$ and $p_{E(D|D>t)}$ denote LACK packet loss probability for both of above cases respectively. That is why call quality „gain" equals

$$\Delta MOS_X(t) = \alpha \cdot \exp(\beta \cdot (p_N(t) + p_{E(D|D>t)}(t))) - \alpha \cdot \exp(\beta \cdot (p_N(t) + p_{E(D)})) \qquad (4\text{-}21)$$

Because probabilities $p_{E(D)}$ and $p_{E(D|D>t)}$ can be expressed as follows

$$p_{E(D)} = \frac{IR(0)}{N_P \cdot P_P} \qquad p_{E(D|D>t)}(t) = \frac{IR(t)}{N_P \cdot P_P} \qquad (4\text{-}22)$$

thus

$$\Delta MOS_X(t) = \alpha \cdot \exp\left(\beta \cdot \left(p_N(t) + \frac{IR(0)}{N_P \cdot P_P}\right)\right) \cdot \left(\exp\left(\frac{-\beta \cdot X(t)}{N_P \cdot P_P}\right) - 1\right) \qquad (4\text{-}23)$$

### 5.3 Dependence of *IR(t)* on estimated probability of the remaining time of the call

Adjusting *IR(t)* based on estimated probability of the remaining time of the call is a proposition of the new hidden data insertion procedure for LACK that has been never considered before.

In previous subsection we considered problem of adjusting *IR(t)* based on estimated average call duration *E(D|D>t)*. In this section we describe adjusting *IR(t)* based on *P(D>T|D>t)* i.e. probability that the call will last longer than *T* under the condition that it already has lasted to $t \le T$ :

$$P(D > T \mid D > t) = \frac{\overline{F}_D(T)}{\overline{F}_D(t)} \qquad (4\text{-}24)$$

Hereafter we analyse dependence of *IR(t)* on *T* value which results from fulfilling the condition $P(D>T|D>t) \ge \xi$, for given *t* from range $[0, \infty)$ and $\xi$ from range $[0, 1]$. For considered in this paper Weibull probability distributions it is equal

$$P(D > T \mid D > t) = e^{\frac{-T^k + t^k}{\lambda^k}} \qquad (4\text{-}25)$$

thus

$$T_\xi(t) \le \sqrt[k]{t^k - \lambda^k \ln \xi} \qquad (4\text{-}26)$$

If the remaining hidden data left to be sent at moment *t* is $S_R(t)$ then

$$\begin{aligned} IR(t) &= \frac{S_R(t)}{T_\xi(t) - t} \ge \frac{S_R(t)}{\sqrt[k]{t^k - \lambda^k \ln \xi} - t}, \quad for\ IR(t) < IR_Q(t) \\ IR(t) &= IR_Q(t), \quad for\ IR(t) \ge IR_Q(t) \end{aligned} \qquad (4\text{-}27)$$

Fig. 16-18 illustrate *IR(t)* curves for Weibull distributions for chosen $C_V$ values, chosen ξ and *S = 1000* bits of steganogram. We assumed that $IR(t)<IR_Q$. The problem related to limiting *IR(t)* by $IR_Q(t)$ is analogous as in previous subsection (see Fig. 11) and so is the solution.

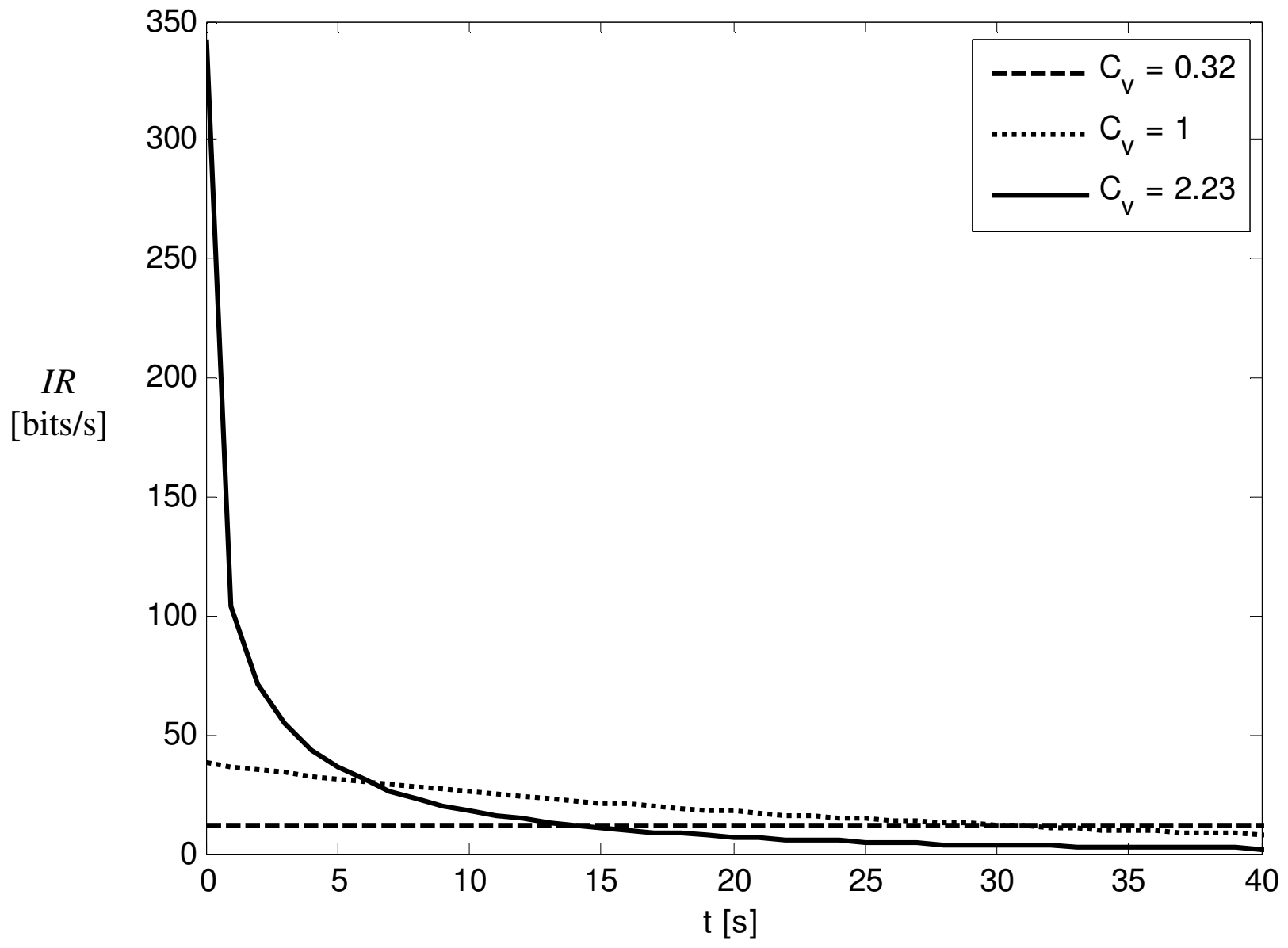

Fig. 16 *IR(t)* for chosen $C_V$ values and $\xi = 0.8$

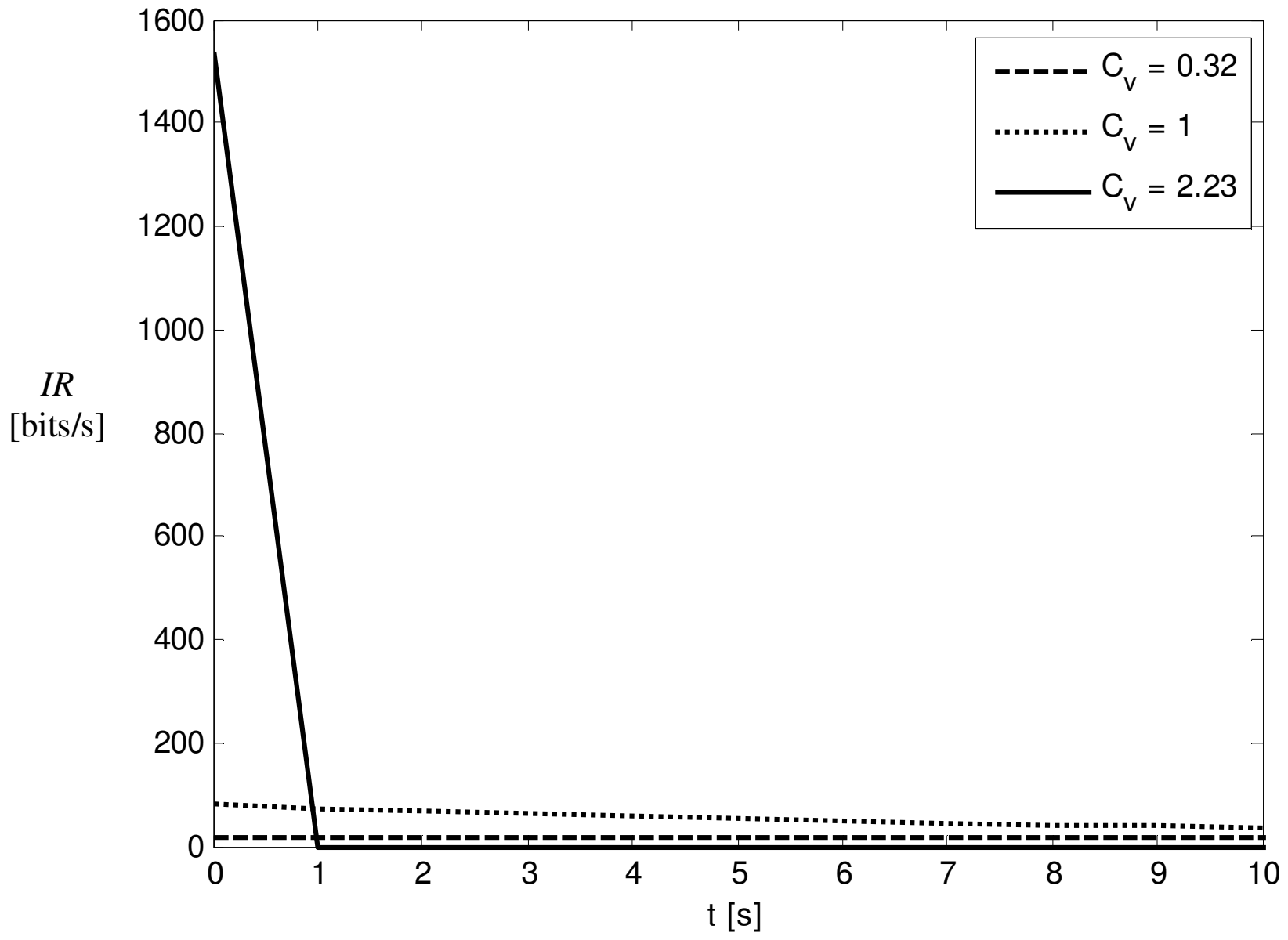

Fig. 17 *IR(t)* for chosen $C_V$ values and $\xi = 0.9$

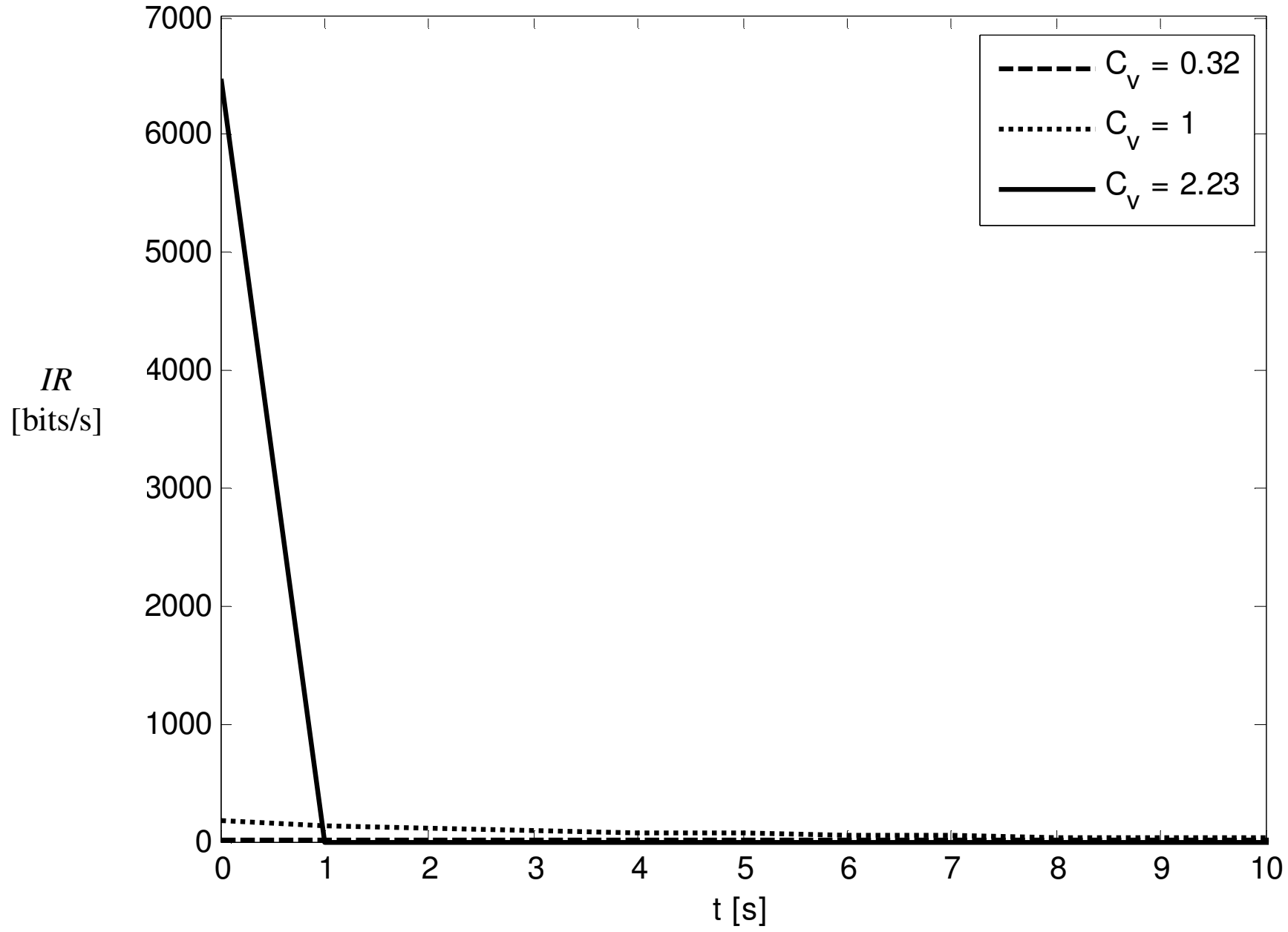

Fig. 18 *IR(t)* for chosen $C_V$ values and $\xi = 0.95$

Fig. 19-21 present dependence of $T_\xi(t)$ for Weibull distribution and chosen values of $C_V$ and $\xi$.

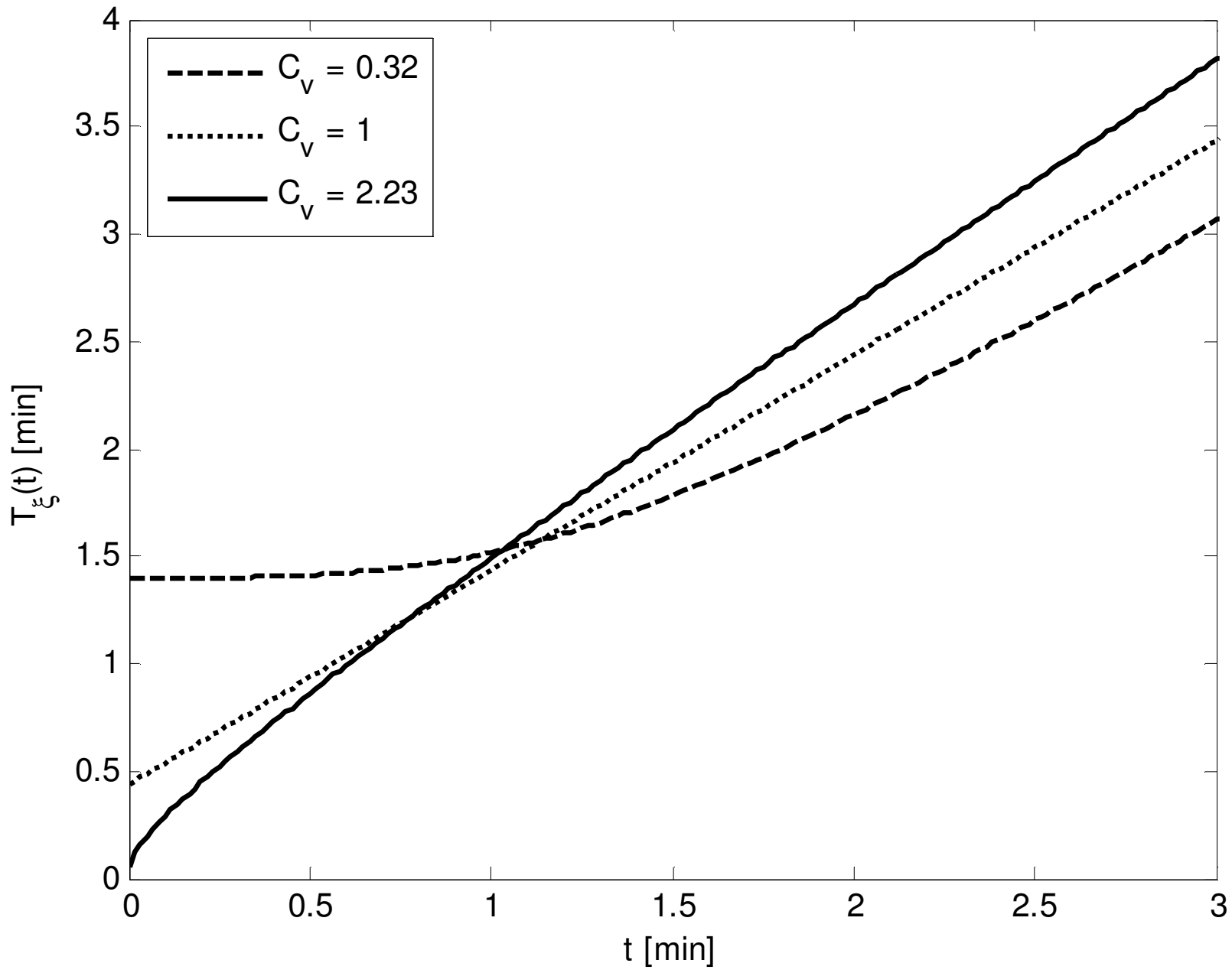

Fig. 19 Dependence of $T_\xi(t)$ on *t* for chosen $C_V = 0.32, 1$ and $2.23$, $\xi = 0.8$

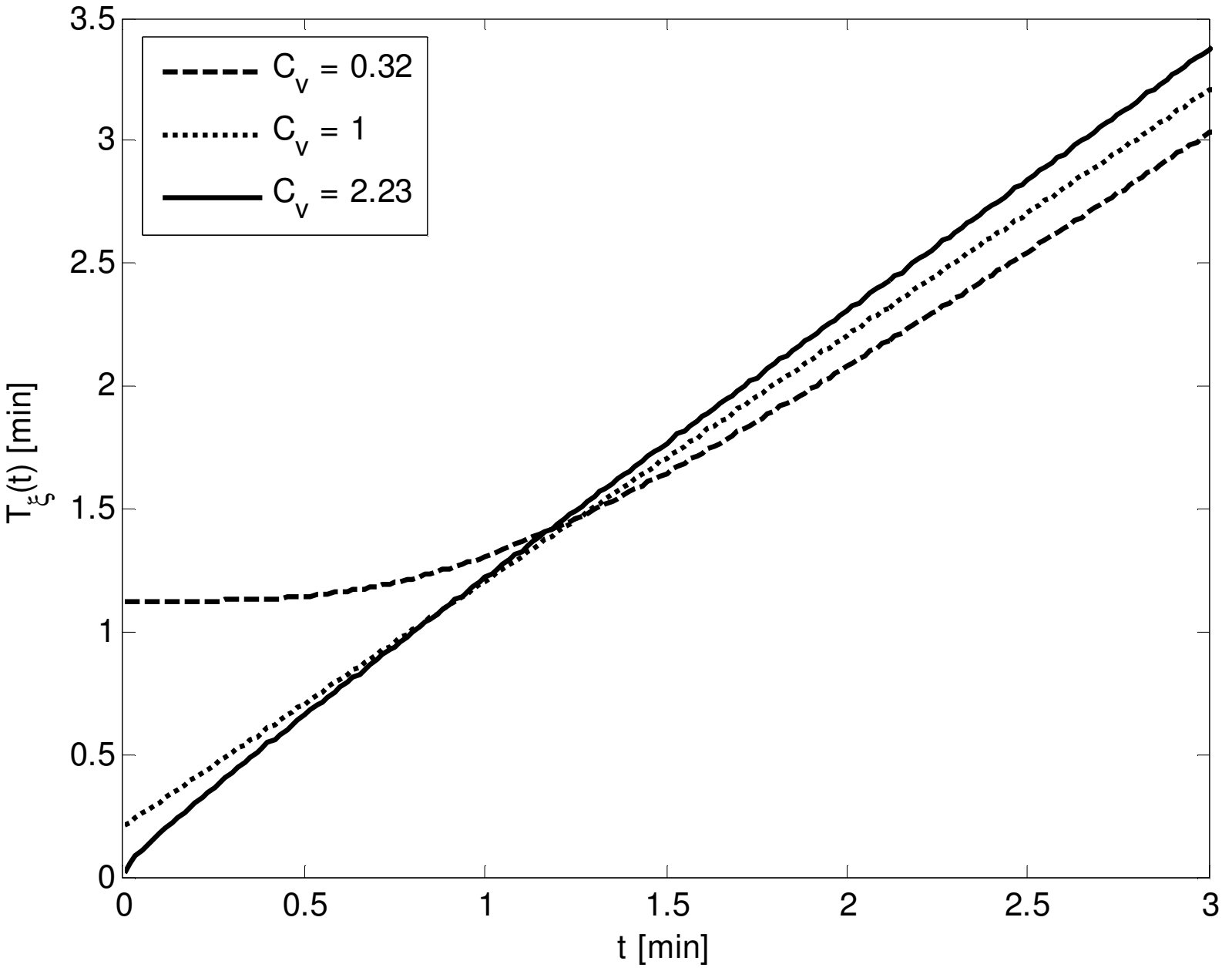

Fig. 20 Dependence of $T_\xi(t)$ on $t$ for chosen $C_V = 0.32, 1$ and $2.23$, $\xi = 0.9$

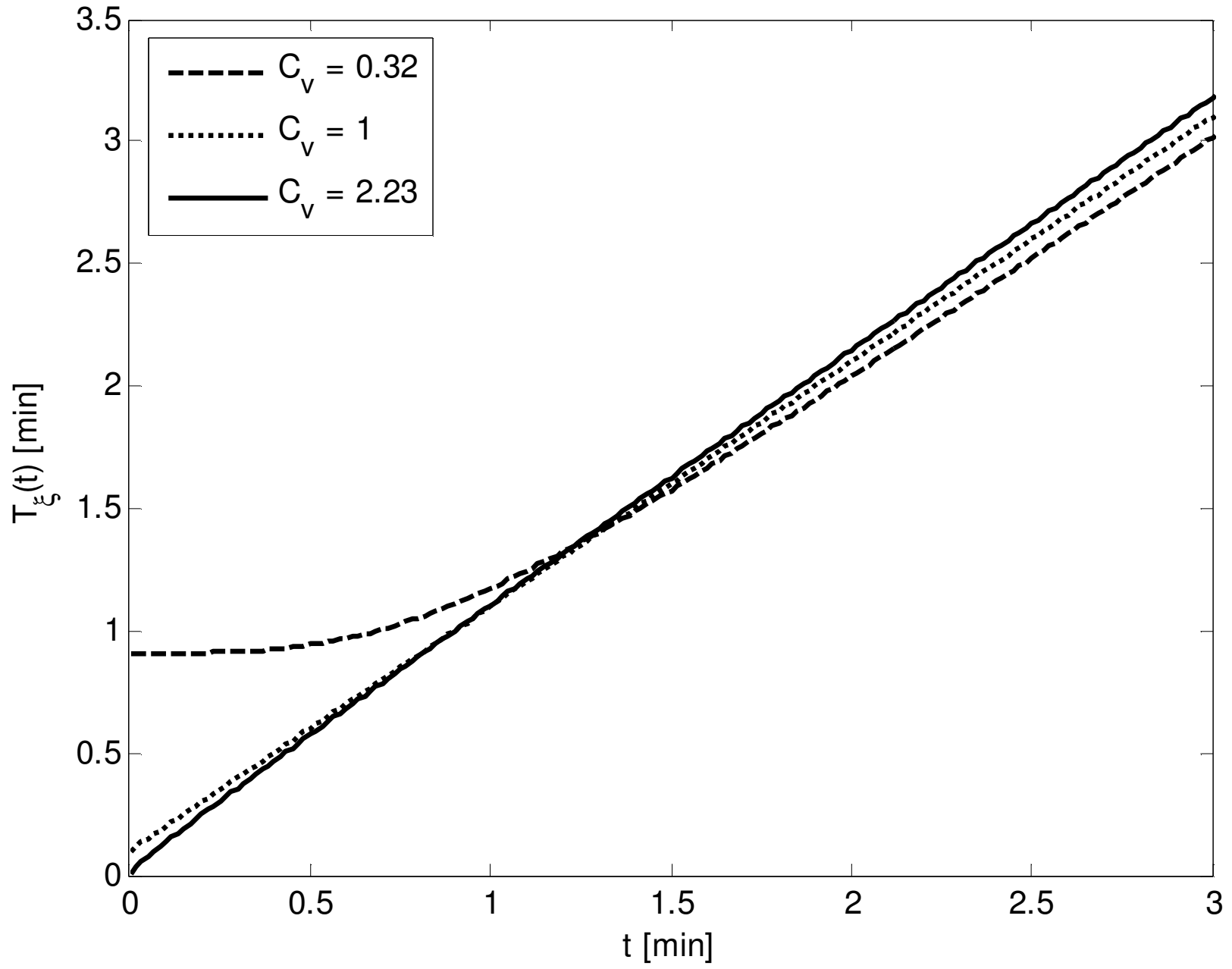

Fig. 21 Dependence of $T_\xi(t)$ on $t$ for chosen $C_V = 0.32, 1$ and $2.23$, $\xi = 0.95$

The curves from Fig. 19 can be approximated with good accuracy as follows

$$T_\xi(t) \approx -0.06\,C_V^2 + C_V\,(0.05\,t + 0.32) + 0.95\,t + 0.17 \qquad (4\text{-}28)$$

Analogous approximations can be achieved for other ξ values.

In Fig. 22-24 dependence of $IR(t)$ on steganogram size for given moments of call is presented under assumption $IR(t) < IR_Q(t)$. For obtained results we assumed the same probability distributions and their parameters as in the previous calculations.

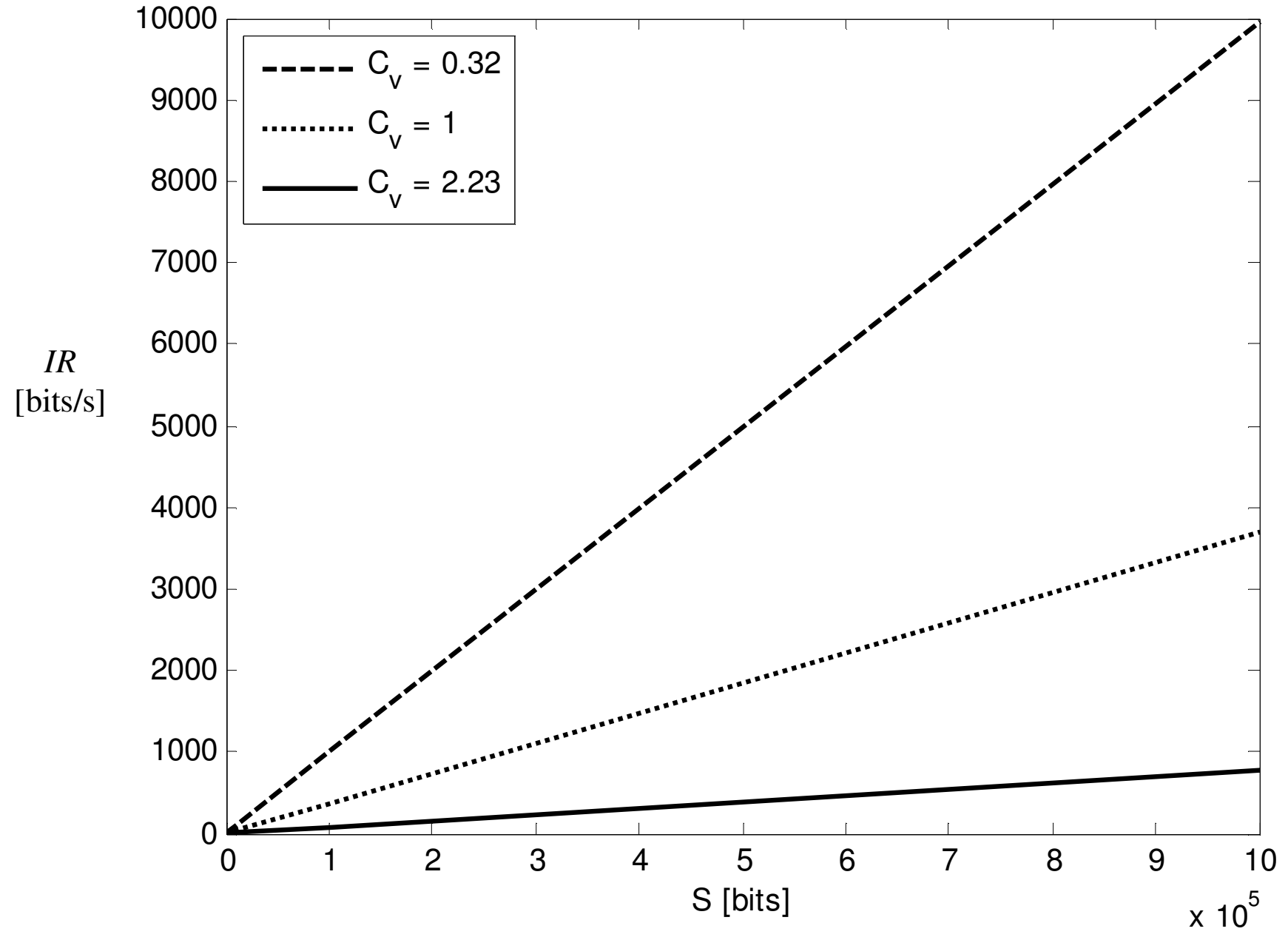

Fig. 22 Dependence of *IR(t)* on *S*, for $t = 60\ s$ and chosen $C_V$ values

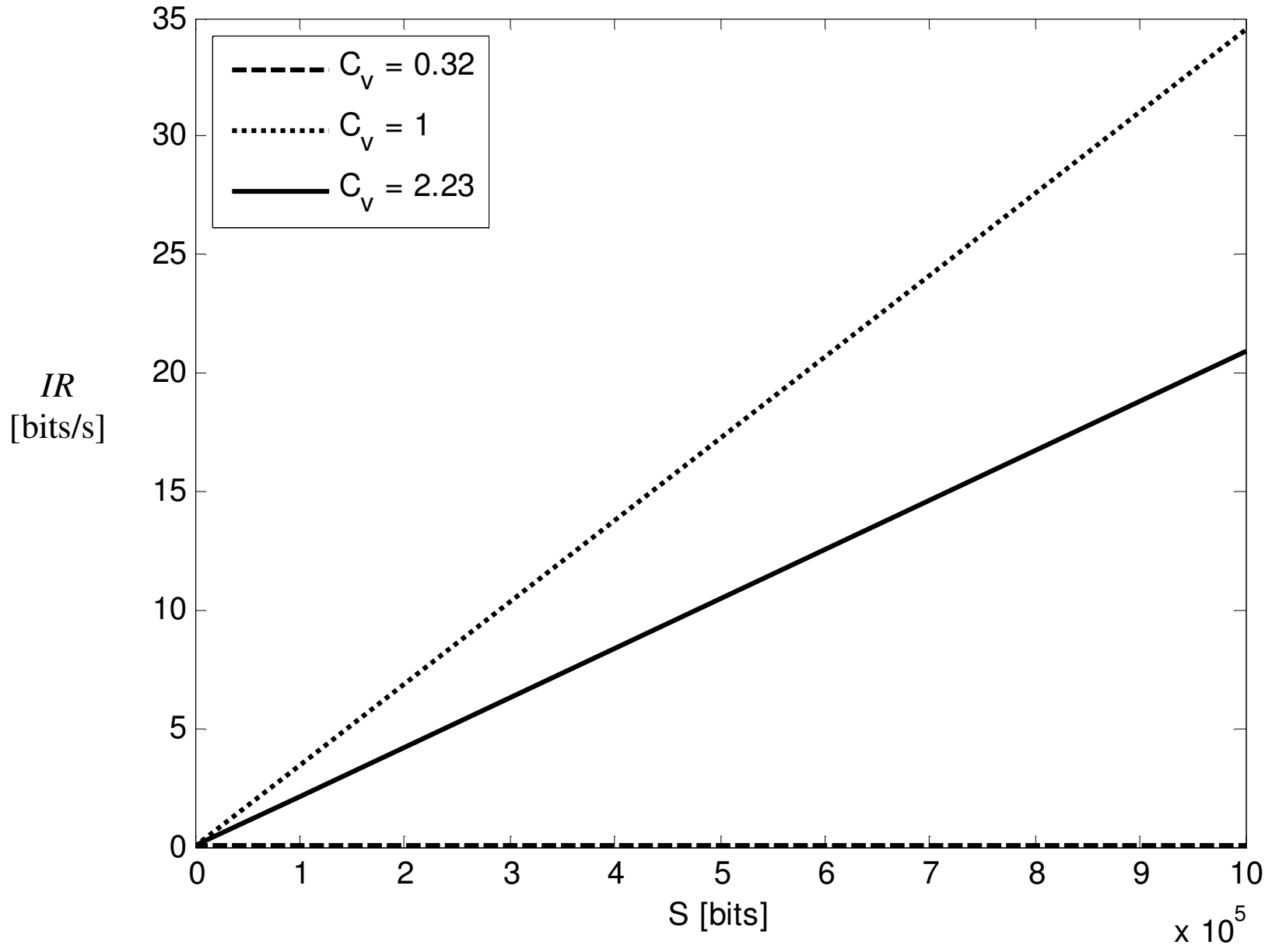

Fig. 23 Dependence of *IR(t)* on *S*, for $t = 180\ s$ and chosen $C_V$ values

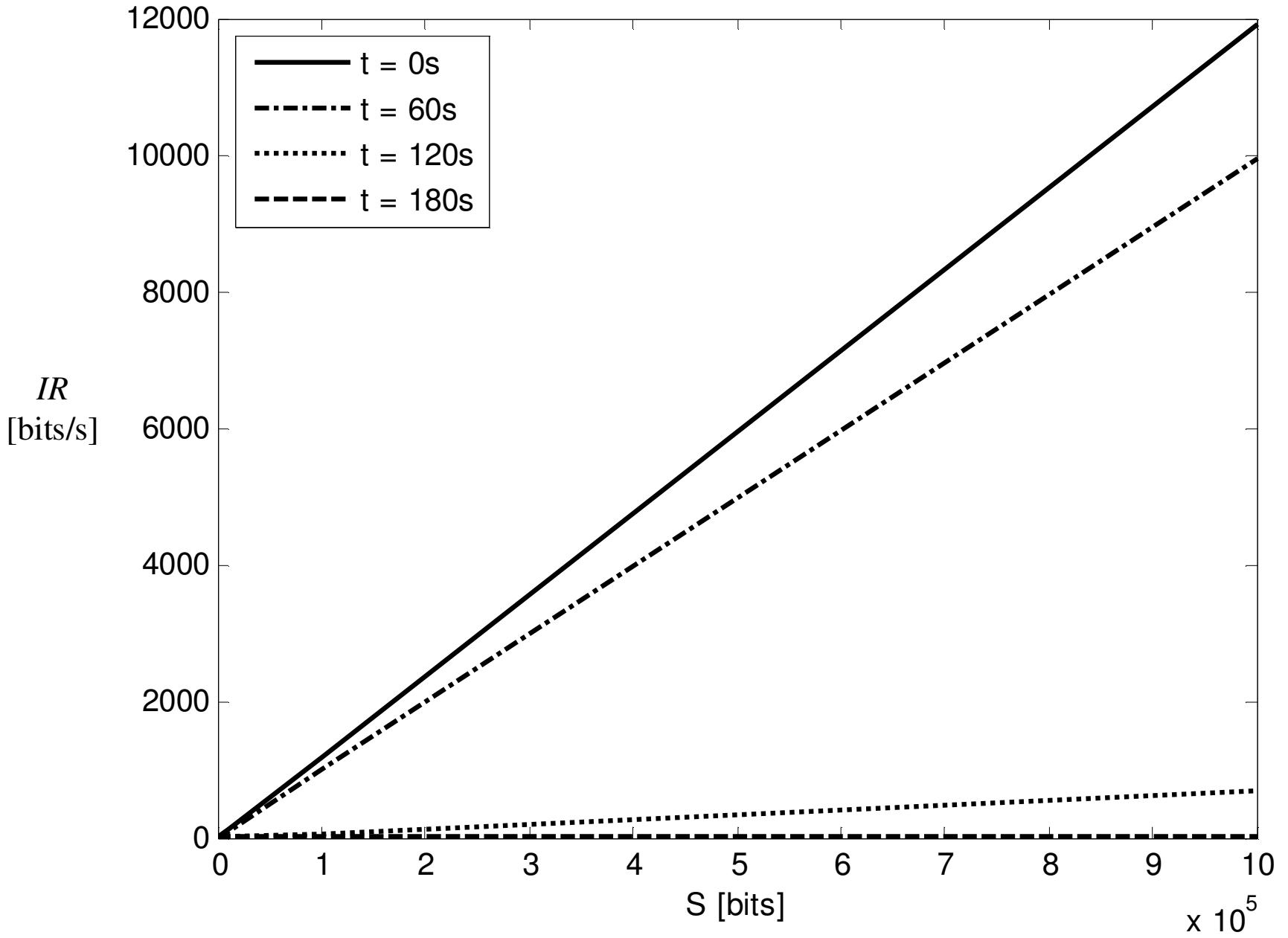

Fig. 24 Dependence of *IR(t)* on *S*, for chosen moments of call and $C_V = 0.32$

### 5.4 Comparison of the methods for adjusting *IR(t)* based on *E(D|D>t)* and *P(D>T|D>t)*

In Fig. 25-27 comparison of methods for adjusting *IR(t)* for both methods presented in subsections 4.2 (based on *E(D|D>t)*) and 4.3 (based on *P(D>T|D>t))* are presented for chosen parameters: $S = 1000$, $C_V = 0.32, 1$ i $2.23$ and $\xi = 0.8$, $0.9$ i $0.95$. To simplify the comparison we assumed $IR(t) < IR_Q(t)$ thus no limitations related to call quality.

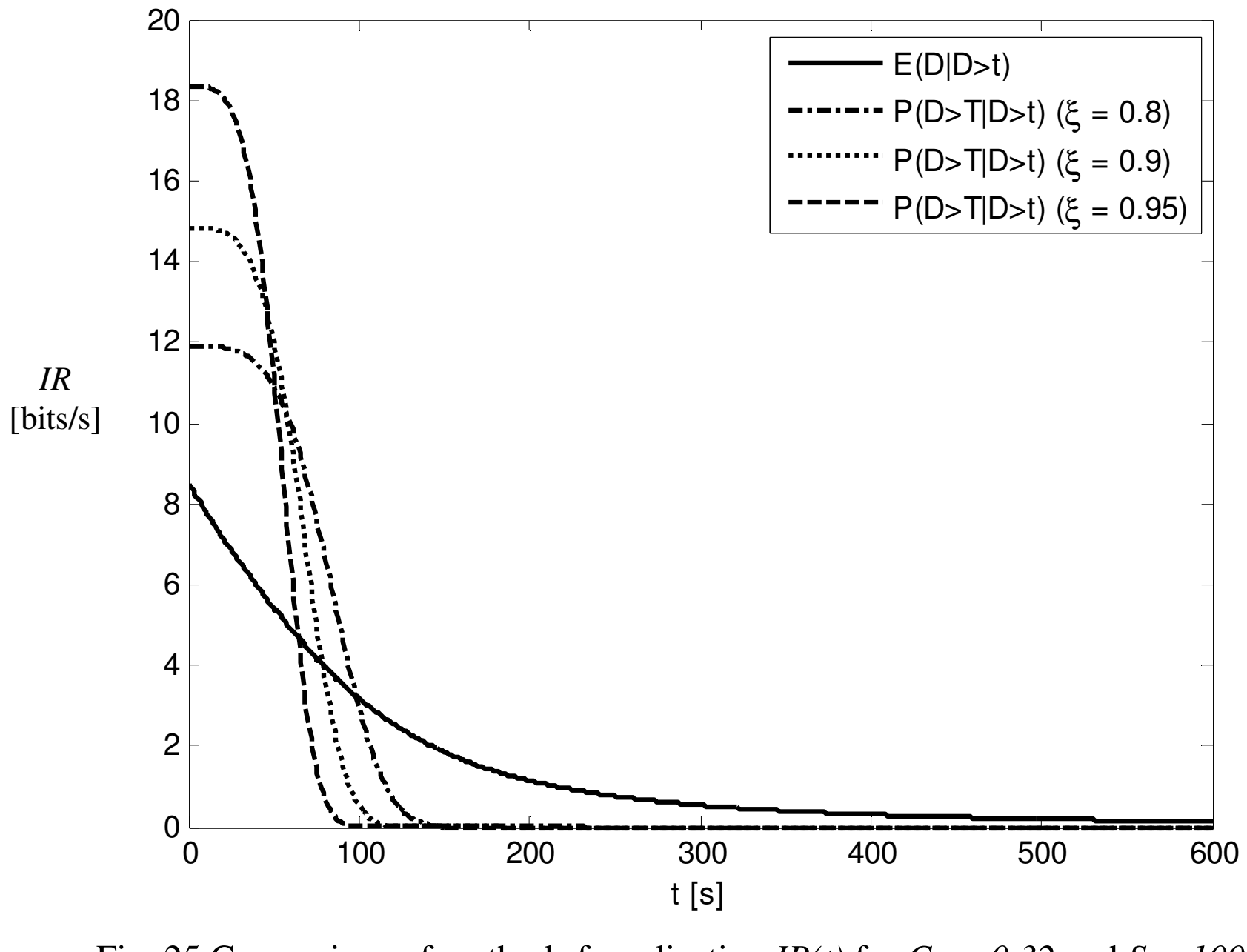

Fig. 25 Comparison of methods for adjusting *IR(t)* for $C_V = 0.32$ and $S = 1000$ bits

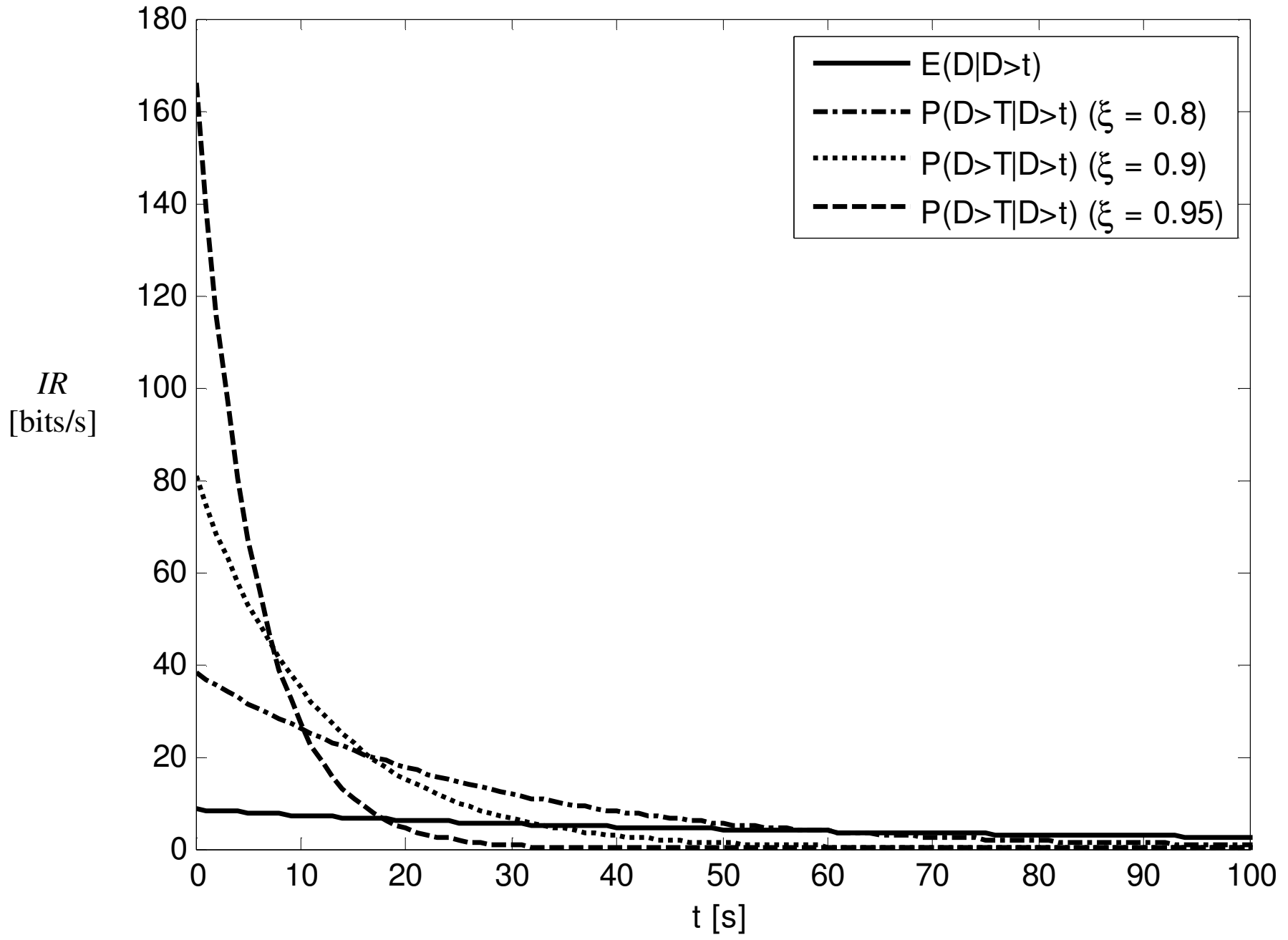

Fig. 26 Comparison of methods for adjusting *IR(t)* for $C_V = 1$ and $S = 1000$ bits

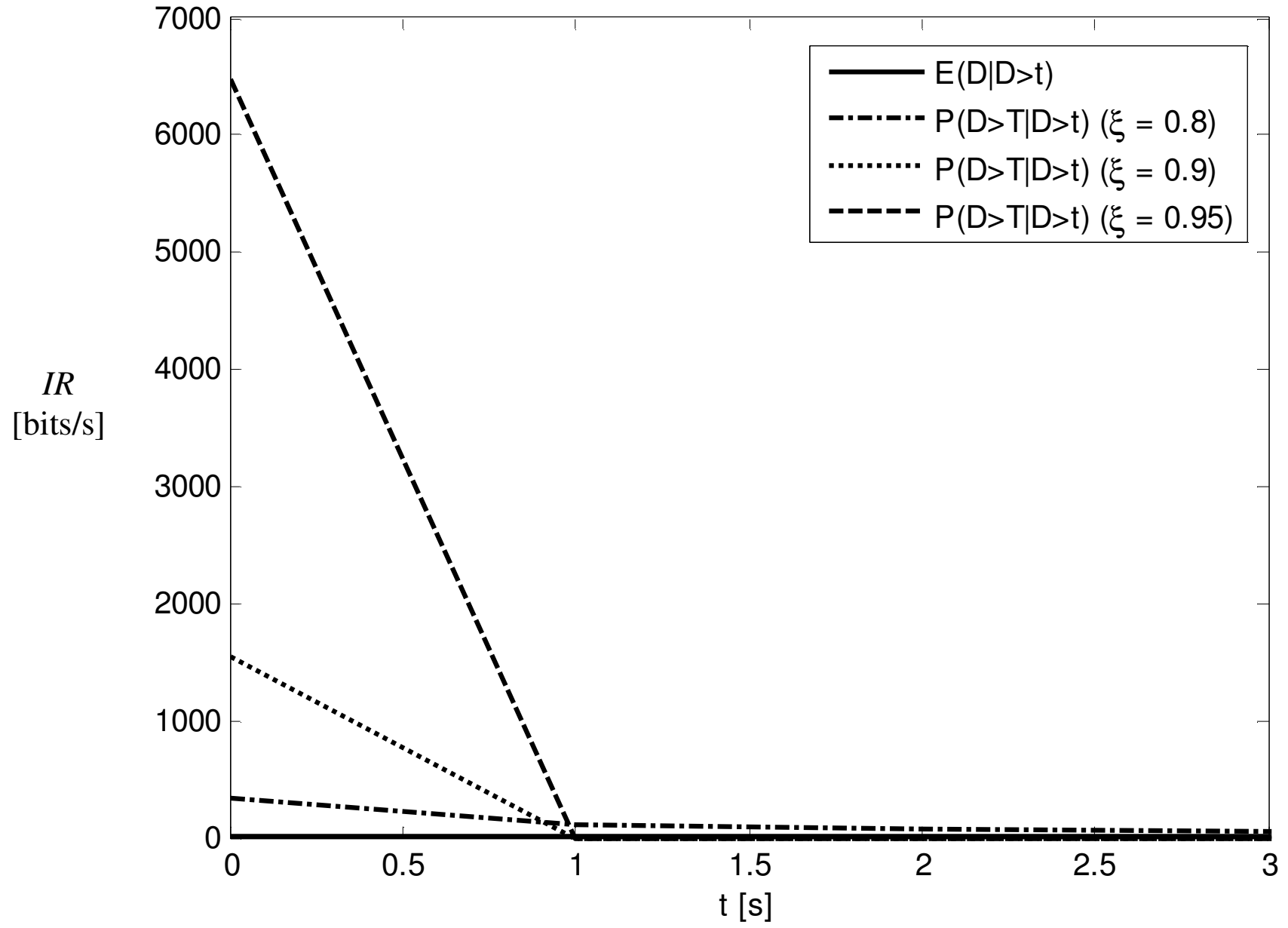

Fig. 27 Comparison of methods for adjusting *IR(t)* for $C_V = 2.23$ and $S = 1000$ bits

Based on figures presented above and analyses carried out in previous subsection we can formulate the following conclusions. Let $IR_{E(D|D>t)}(t)$ and $IR_{P(D>T|D>t)}(t)$ denote hidden data insertion rates for method based on $E(D|D>t)$ and $P(D>T|D>t)$ respectively.

For the beginning of the call $IR_{E(D|D>t)}(t) \leq IR_{P(D>T|D>t)}(t)$ ($t \leq t'$, and depends mainly on $C_V$). If $IR_Q(t) \leq IR_{E(D|D>t)}(t)$in range $[0, t')$ for both methods we are witnessing hidden data insertion "arrear" and it is smaller for the method based on $E(D|D>t)$. This "arrear" must be aligned later during the call after the moment $t'$, so it requires increasing $IR_{E(D|D>t)}(t)$ and $IR_{P(D>T|D>t)}(t)$, for $t > t'$. However, the degree of increasing $IR_{E(D|D>t)}(t)$ is smaller than for $IR_{P(D>T|D>t)}(t)$ which is beneficial from the call quality and resistance to steganalysis point of view (if $IR_{P(D>T|D>t)}(t) \geq IR_J(t) \geq IR_{E(D|D>t)}(t)$ in range $[0, t')$, then method based on $E(D|D>t)$ does not introduce any "arrear").

In time intervals in which $IR_{P(D>T|D>t)}(t) \geq IR_{E(D|D>t)}(t)$, the method based on $E(D|D>t)$ potentially has lower negative influence on call quality and resistance to steganalysis. On the other hand, in the time intervals in which $IR_{P(D>T|D>t)}(t) \leq IR_{E(D|D>t)}(t)$ method based on $P(D>T|D>t)$ is, for the same reasons, potentially more valuable.

The greater $IR_{P(D>T|D>t)}(t)$ and $IR_{E(D|D>t)}(t)$ the potentially greater steganographic bandwidth. Thus, from this point of view more favourable is the method, for given time intervals, for which hidden data insertion rate is greater. That is why, if we consider LACK call quality and resistance to detection it is more rational to utilise the method for adjusting $IR(t)$ based on $E(D|D>t)$. Whereas, if we consider LACK steganographic bandwidth then more advantageous is method based on $P(D>T|D>t)$.

Thus, the choice of the method for adjusting $IR(t)$ requires making a trade-off between desired call quality, resistance to steganalysis and desired steganographic bandwidth. This trade-off depends on the context and application of LACK and that is why it cannot be established arbitrarily.

One must always take under consideration that mutual relationships between presented methods depend mainly on statistical properties of VoIP call duration and on $C_V$ in particular. If we acknowledge that presented experimental data (see Section 4.1) is representative for IP telephony, at least when it comes to average and variance of the call duration, then only $C_V$ substantially greater than 1 should be considered. Thus, mutual relationships between $IR_{P(D>T|D>t)}(t)$ and $IR_{E(D|D>t)}(t)$ will be similar to those presented in Fig. 27.

## 6. Conclusions and Future Work

In this paper LACK steganographic method was subjected to the detailed performance evaluation. We have focused on two hidden data insertion rate *IR* procedures (first: based on estimated remaining average call duration and second: based on the estimated probability of the remaining time of the call) and their dependence on estimated call duration and voice quality.

It was shown that the insertion rate may be effectively made dependent on the current call duration time, and that this dependence can be expressed with good accuracy with the coefficient of variation of the call duration probability distribution. We have also derived analytical relations which enable making $IR(t)$ dependent on voice quality parameters. All derived formulae are simple and can be straightforwardly implemented. Comparison of the both presented procedures was also included. It showed that the choice of the method for adjusting $IR(t)$ requires making a trade-off between desired call quality, resistance to steganalysis and desired steganographic bandwidth.

The effectiveness of the resulting hidden data insertion procedures will depend on the accuracy of the estimated mean call duration, the coefficient of variation of the call duration and the probability distribution of voice quality for the network (sub-network), which is intended to be used for sending steganographic data with the LACK method. Thus to evaluate realistically this effectiveness more experimental data has to gathered, nevertheless the authors believe that the analysis presented in this paper indicates that LACK provides good chance for high effectiveness.

Future work will include conducting experiments for LACK in real VoIP network and assessing the practical steganographic bandwidth and resistance to detection for different network conditions, types of jitter buffers and voice codecs that can be achieved without excessively degrading the call quality.

## ACKNOWLEDGMENTS

- This work was partially supported by the Polish Ministry of Science and Higher Education under Grant: N517 071637.
- The authors would like to thank:
  - R. Birke, M. Mellia, M. Petracca and D. Rossi from Politecnico di Torino (Italy) for sharing details of their VoIP experimental data
  - Shiguo Lian from France Telecom R&D, Beijing (China) for valuable comments and fruitful discussions.